\documentclass[12pt,preprint]{aastex}

\slugcomment{}
\shortauthors{Fujii et al.}
\shorttitle{Colors of a Second Earth}
\begin{document}
\title{Colors of a Second Earth:\\
Estimating the fractional areas of ocean, land, and vegetation\\
of Earth-like exoplanets}
\author{
 Yuka Fujii\altaffilmark{1}, 
 Hajime Kawahara\altaffilmark{1},
 Yasushi Suto\altaffilmark{1,2,3}, 
 Atsushi Taruya\altaffilmark{1,2,4},
 Satoru Fukuda\altaffilmark{5}, 
 Teruyuki Nakajima\altaffilmark{5}, and
 Edwin L.~Turner \altaffilmark{3,4}}  

\altaffiltext{1}{Department of Physics, The University of Tokyo, 
Tokyo 113-0033, Japan}
\altaffiltext{2}{Research Center for the Early Universe, Graduate School
of Sciences, The University of Tokyo, Tokyo 113-0033, Japan}
\altaffiltext{3}{Department of Astrophysical Sciences, 
    Princeton University, Princeton, NJ 08544, USA}
\altaffiltext{4}{Institute for the Physics and Mathematics of the
Universe, The University of Tokyo, Kashiwa 277-8568, Japan}
\altaffiltext{5}{Center of Climate System Research, 
The University of Tokyo, Kashiwa 277-8568, Japan}
\email{yuka.fujii@utap.phys.s.u-tokyo.ac.jp}
\begin{abstract}
Characterizing the surfaces of rocky exoplanets via their scattered light
will be an essential challenge in investigating their habitability and 
the possible existence of life on their surfaces. 
We present a reconstruction method for fractional areas of different 
surface types from the colors of an Earth-like exoplanet. 
We create mock light curves for Earth without clouds using empirical data. 
These light curves are fitted to an isotropic scattering model consisting 
of four surface types: ocean, soil, snow, and vegetation. 
In an idealized situation where the photometric errors are only photon shot 
noise, we are able to reproduce the fractional areas of those components 
fairly well. 
The results offer some hope for detection of vegetation via the distinct 
spectral feature of photosynthesis on the Earth, known as the red edge. 
In our reconstruction method, Rayleigh scattering due to the atmosphere 
plays an important role, and for terrestrial exoplanets with an atmosphere 
similar to our Earth, it is possible to estimate the presence of oceans 
and an atmosphere simultaneously.
\end{abstract}

\section{introduction}
\label{s:intro}

A major milestone in exoplanet research will be the discovery of
Earth-like planets. Indeed planets a few times as massive as the Earth,
{\it super-earths} have already been discovered using the radial velocity
method; for instance, $M_{\rm p} \sin i$ of GJ581e is estimated to 
$\sim 2M_{\oplus}$ where $M_{\rm p}$ is the planetary mass and $i$ is 
the inclination of its orbit \citep{mayor2009}. 
Microlensing and transit methods are well-suited for the detection of 
such low-mass planets; MOA-2007-BLG-192-Lb \citep{bennett2008}
and CoRoT-7b \citep{queloz2009} are reported to have masses of $3.3
{+4.9 \atop -1.6} M_{\oplus}$ and $4.8 \pm 0.8 M_{\oplus}$,
respectively.  The {\it Kepler} mission, launched on 2009 March 6,
aims to detect a number of terrestrial planets with masses 
on the order of $M_{\oplus}$. 
Such low-mass planets are very likely to be rocky
and may have bodies of liquid water on their surfaces if they orbit in 
Habitable Zone \citep[HZ, e.g.,][]{kasting1993,kasting2003} of the primary 
star. Their discovery will definitely trigger ever more serious investigations 
of techniques to search for signatures of life, or {\it biomarkers}.

The most conventional and extensively studied biomarkers are based on
spectroscopic identification of molecular species in the planetary
atmospheres, such as ${\rm O}_2$ and ${\rm O}_3$ \citep[e.g.,][]{leger1993, desmarais2002}.  For the most favorable known
exoplanetary systems, detection of such atmospheric absorption features is
already within reach during the transit or the secondary eclipse;
absorption features of ${\rm Na}$, ${\rm C}$, ${\rm O}$, and ${\rm H}$ in
a hot Jupiter HD209458b \citep{charbonneau2002, vidal2003, vidal2004,
ballester2007} and ${\rm H}_2{\rm O}, {\rm CH}_4$, ${\rm CO}$, and ${\rm
CO}_2$ in HD189733b \citep{tinetti2007, swain2008, swain2009} have been
reported. Spectroscopy of non-transiting terrestrial planets orbiting in
the HZ will likely require major and specialized observatories in space,
but such facilities are being actively studied and evaluated by both
NASA and ESA at present \citep[e.g.,][]{tpf-c, tpf-i, darwin}.

An even more ambitious and direct approach to the search for life may be
possible via multi-band photometry of terrestrial exoplanets that is 
observationally less demanding.  In particular, the light
scattered by the surface of a planet carries important information on
properties of the surface.  \citet{ford2001} computed for the first time
the diurnal variation of scattered light by the Earth observed at a
distance of 10 pc.  They showed that the fractional variation of
light curves of the Earth is 10 \% -20\%, which indeed agrees with the
result of Earthshine observations \citep[e.g.,][]{goode2001}.

More importantly, these model light curves exhibited an increase in
brightness due to the presence of vegetation around at $\lambda =
$750 nm.  This sharp increase of the albedo is known as the {\it red
edge}. It is a fairly generic feature of vegetation on the Earth and is
due to bio-pigments associated with photosynthesis. The red edge feature
has been detected by Earthshine observations
\citep{woolf2002,arnold2002,rodriguez2006}. It was subsequently discussed 
as a possible biomarker in extrasolar terrestrial planets by
\citet{seager2005} and \citet{kiang2007a,kiang2007b}.
\citet{tinetti2006a, tinetti2006b} and \citet{rodriguez2006} performed
more accurate simulations of the scattered radiation spectra of the
Earth. They paid particular attention to the red edge, and discussed the
detectability of its spectral feature.

\citet{palle2008} focused on the determination of the rotation
period of an extrasolar terrestrial planet from the time variation of
its scattered light. They concluded that the period can be reliably 
estimated in the presence of realistic partial cloud coverage of the 
planetary surface. They even found that the global circulation of the 
cloud pattern systematically modulates the estimate of the spin 
rotation period of the planet.

More recently, \citet{cowan2009} approached the problem in a different
and interesting manner. They performed a principal component analysis
(PCA) on the real light curves of the Earth observed by the Extrasolar 
Planet Observation and Deep Impact Extended Investigation (EPOXI) 
mission. They identified two major eigenspectra that roughly
correspond to ocean and land on the Earth and extracted a rough
distribution of these components as a function of longitude.  This was
the first attempt to solve the inverse problem and to extract the
surface properties from the observed data in a model-independent
fashion.  \citet{williams2008} and \citet{oakley2009} paid attention to
the variation of scattered light according to the orbital motion of the
planet. They pointed out that the reflectivity of ocean dramatically
increases at a crescent phase and that the existence of the liquid water
may be observationally inferred through the effect.  \citet{oakley2009}
also suggested a possibility of reconstructing the land/ocean
distribution along longitude using the gap between the reflectivity of
ocean and that of land.

In this paper, we develop still another method to estimate the fractional 
areas of different surface types on extrasolar planets from multi-band
photometry. 
We are particularly interested in terrestrial exoplanets in habitable zone 
and consider the detectability of vegetation using the red-edge feature. 
Our attempt is to reproduce the scattered light curves as a sum of the 
four surface types, i.e., ocean, soil, vegetation, and snow.  While PCA 
attempts to extract only orthogonal eigenspectra in a model-independent 
manner, the correspondence with real surface types is not clear.  
Indeed, we would like to answer an ambitious, but well-defined, scientific 
question: ``If we discover an Earth-like exoplanet in the near future, 
to what extent can we reconstruct the surface information, in particular 
the presence of vegetation observationally?''. For that purpose, we 
intentionally and specifically adopt the major surface types of the Earth, 
including vegetation, into the analysis even though our method is
thus inevitably model dependent.  We discuss the feasibility of our
method by applying it to mock scattered light curves of the {\it cloudless}
Earth. We hope that our current results are relevant for a future TPF
mission \citep{tpf-c} such as the Occulting Ozone Observatory which aims 
at multi-band photometry of Earth-like planets \citep{kasdin2010}.

The rest of this paper is organized as follows. Section
\ref{s:simulation} describes the generation of mock light curves of the
Earth. The methodology, assumptions, and results of our inversion
model to reconstruct the fractional areas are presented in Section 
\ref{s:inverse}.  We discuss the importance of the photometric band
selection and comparison with PCA in Section \ref{s:dis}.  Finally we
summarize the conclusion and discuss future directions of investigation
in Section \ref{s:sum}.

\section{Mock scattered light curves of the Earth}
\label{s:simulation}

\subsection{Configuration of the Star-planet-observer System}
\label{ss:config}

\begin{figure}[!h]
  \centerline{\includegraphics[width=13.0cm]{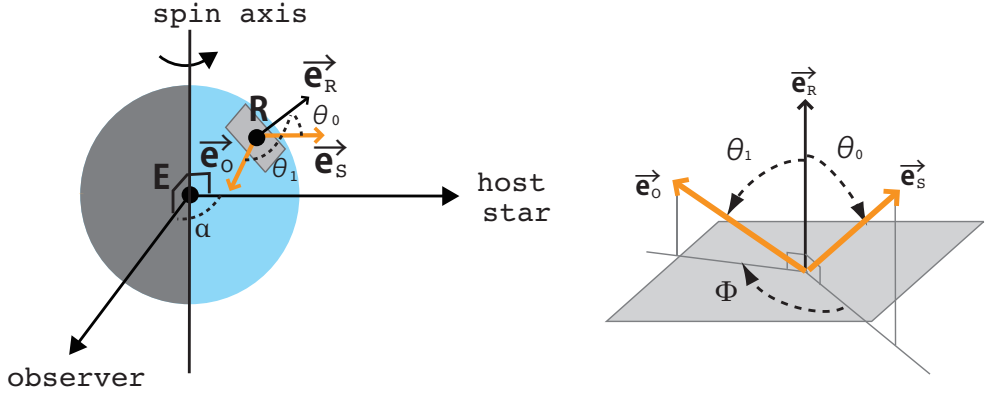}}
\caption{Schematic illustration of our fiducial configuration of the
planetary system.  The star is located at the equinox and the observer
is in the equatorial plane.  The phase angle $\alpha $ denotes the
planet-centric angle between the host star and the observer, which is
fixed to $\alpha = \pi/2$.  The unit vectors $\protect{\bf e}_{\rm S}$
and $\protect{\bf e}_{\rm O}$ go from the center of the planet (E)
toward the host star (S) and the observer (O), respectively.  The vector
$\protect{\bf e}_{\rm R}$ is a unit vector normal to the surface plane
at R. The solar zenith angle $\theta _0$, the zenith angle of the
observation $\theta _1$, and the relative azimuthal angle $\phi $ are
defined at each point R as the left panel indicates. As mentioned in 
Section \ref{s:model}, the illuminated pixels satisfy $\protect{\bf e}_{\rm S}
\cdot \protect{\bf e}_{\rm R} > 0$, and visible pixels satisfy
$\protect{\bf e}_{\rm O} \cdot \protect{\bf e}_{\rm R} > 0$.}
\label{fig:config}
\end{figure}

This section describes our computation method for generating mock light
curves that are analyzed in the following section.  Figure \ref{fig:config}
illustrates the geometric configuration of the star-planet system
adopted in this paper. 
The vectors ${\bf e}_{\rm S}$, ${\bf e}_{\rm R}$, and ${\bf e}_{\rm O}$
denote the unit vectors from the center of the planet (E) toward the
host star (S), an arbitrary point on the surface (R), and the observer
(O), respectively.  
A phase angle, $\alpha $, is defined as an angle
between ${\bf e}_{\rm S}$ and ${\bf e}_{\rm O}$, which is fixed to
$\alpha = \pi /2$ in this paper, except where noted otherwise.  
The solar zenith angle, $\theta _0 $,
at R is an angle between ${\bf e}_{\rm S}$ and ${\bf e}_{\rm R}$, and
the zenith angle of the observation, $\theta _1$, is that between ${\bf
e}_{\rm R}$ and ${\bf e}_{\rm O}$. 
The azimuthal angle between incident direction and the direction of 
observation is represented by $\phi$ as 
shown in the right panel of Figure \ref{fig:config}.  In our fiducial
configuration, the planet is located at the equinox (except in Section 
\ref{s:integrate}) and the observer is on the equatorial plane.

The incident ray from the host star toward the planet first passes
through the atmosphere, which may scatter the ray. A fraction of the
light reaches the planetary surface and is scattered there, and then
returns to space passing through the scattering atmosphere once again.
The scattering properties of the atmosphere depend on its composition
and overall optical thickness, while those of the
surface vary according to the details such as components, coarseness,
moisture, the shape of surface, and so on.  The wavelength-dependence of
the synthetic scattered light therefore carries information about the
atmosphere as well as the surface features.  The fact that the illuminated
and visible part of the planet gradually changes in time due to the
motion of the planet (orbiting around its host star and spinning about
its own axis) allows observers to scan the planetary surface in a
fashion similar to global remote sensing of the Earth observation
satellites.  Thus, the variation of planetary light curves in different
photometric bands allows mapping the
planetary surface to some extent.

\subsection{Model and Assumptions}
\label{s:model}

To generate synthetic scattered light curves, we approximate the
planetary sphere by polygons of  $2^{\circ }.5 \times 2^{\circ }.5$
``pixels.''  The angles $\theta _0$, $\theta _1$, and $\phi$ are defined at
the center of each pixel.  In general, the radiance $L (\theta _0, \theta
_1, \phi ; \lambda )\,({\rm W\,str^{-1}\,m^{-2}\, \mu m^{-1}})$ of each pixel
is expressed as
\begin{equation}
\label{eq:radiance}
L (\theta _0, \theta _1, \phi ; \lambda ) 
= F_* (\lambda ) \cos \theta _0 \, f(\theta _0, \theta _1, \phi ; \lambda ),
\end{equation}
where $F_*(\lambda )\,({\rm W\,m^{-2}\,\mu m}^{-1})$ is the incident
flux at wavelength $\lambda $, and $f(\theta _0, \theta _1, \phi;
\lambda )\,({\rm str^{-1}})$ characterizes the scattering property of
the surface and is conventionally referred to as the bidirectional 
reflectance distribution function (BRDF).  The intensity of the total 
scattered light of a planet, 
$I(\lambda )\,({\rm W\,str^{-1}}\,\mu {\rm m}^{-1})$, is obtained by
integrating $L$ over the illuminated and visible region of the planetary
surface $S$:
\begin{eqnarray}
\label{eq:intensity}
I(\lambda ) &=& \int _S  L(\theta _0, \theta _1, \phi ; \lambda) 
\cos \theta _1 dS \nonumber \\
&=& \int _S F_* (\lambda ) f(\theta _0, \theta _1, \phi ; \lambda) 
\cos \theta _0 \cos \theta _1 dS \nonumber \\
&=&  F_* (\lambda ) R_{\rm p}^2  \int _s 
f(\theta _0, \theta _1, \phi ; \lambda) 
\cos \theta _0 \cos \theta _1 ds,
\end{eqnarray}
where $R_{\rm p} $ is the radius of the planet, $ds \equiv dS/R_{\rm p}^2$ is
a normalized area element, and the illuminated and visible area $S$ is 
defined as a set of pixels that satisfy ${\bf e}_{\rm S}\cdot {\bf e}_{\rm R} \ge 0$ 
and ${\bf e}_{\rm O}\cdot{\bf e}_{\rm R} \ge 0$ simultaneously. 
Since the scattering property of a particular position on the surface 
gradually changes as the planet spins, $I(\lambda )$ varies as a function 
of time.  Equation (\ref{eq:intensity}) is a key formula in computing 
the synthetic scattered light curve.

The evaluation of Equation (\ref{eq:intensity}) requires three inputs: 
$R_{\rm p}$, $F_*(\lambda )$ and $f(\theta _0, \theta _1, \phi ; \lambda
)$. 
We approximate the incident flux $F_*(\lambda )$ from the star by a 
black-body spectrum:
\begin{equation}
\label{eq:F}
F_* (\lambda ) = \frac{2\pi h c^2 R_*^2}{ \lambda ^5}
\frac{1}{\displaystyle \exp \left( hc/\lambda k_B T_* \right)-1} 
\frac{1}{d^2},
\end{equation}
where $h$ is the Planck constant, $c$ is the speed of light, $k_B$ is
the Boltzmann constant, $R_*$ and $T_*$ are the radius and the surface
temperature of the star, and $d$ is the distance between the star and
the planet.  In this paper, we assume the Sun-Earth system, and therefore
we set $R_{\rm p} = R_{\oplus} = 6.4 \times 10^6\;{\rm m}$, 
$T_* = T_{\odot}=5800\;{\rm K}$, $R_*=R_{\odot}=7.0\times 10^8\;{\rm m}$, 
and $d=1\;{\rm AU}$.

We adopt a single scattering approximation for the interaction between
the atmosphere and the underlying surface, i.e., the light is
scattered once at most either in the atmosphere or at the surface 
and all multi-scattering processes are ignored.  Then, the radiance of 
each pixel can be written as a linear combination of those associated 
with the  surface and the atmosphere:
\begin{equation}
L(\theta _0, \theta _1, \phi; \lambda ) 
= L_{{\rm atm}}(\theta _0 , \theta _1, \phi; \lambda ) 
+ L_{{\rm surf}}(\theta _0, \theta _1, \phi;\lambda ) ,
\end{equation}
or equivalently in terms of the BRDF (Equation (\ref{eq:radiance})),
\begin{equation}
f(\theta _0, \theta _1, \phi ; \lambda ) 
= f_{\rm atm}(\theta _0, \theta _1, \phi ; \lambda) 
+ f_{\rm surf}(\theta _0, \theta _1, \phi ; \lambda).
\end{equation}
The specific models for $f_{\rm atm}$ and $f_{\rm surf}$ are described
below.

\subsubsection{$f_{\rm atm}$ --- Scattering by Atmosphere}

For scattering in the atmosphere represented by $f_{\rm atm}$, we consider
Rayleigh scattering alone, and ignore the effects of clouds, aerosols, and
molecular absorption, which will be discussed elsewhere.  We use a 
plane-parallel approximation locally for each pixel because our pixel size of 
$2.5^\circ\times 2.5^\circ$ is sufficiently small that one can reasonably 
approximate the overall sphere by a polygon consisting of such pixels. 
While this approximation is not valid for pixels near the edge of the 
illuminated and visible area, the fractional area of the
region is small and our results are not changed significantly.

Under the single-scattering approximation, the radiance from the
atmosphere, $L_{{\rm atm}}$, is given by
\begin{eqnarray}
\label{eq:Latm}
&& L_{{\rm atm}}(\theta _0, \theta _1, \phi ;\lambda) 
= F_{*}\int _0 ^{\tau } \omega \Psi(\Theta)
\exp \left\{ -\tau ' \left( \frac{1}{\cos \theta  _0} 
+ \frac{1}{\cos \theta _1} \right) \right\} 
\frac{d\tau ' }{\cos \theta _1} \nonumber \\
&& \qquad = \frac{F_* \omega \Psi(\Theta )}{\cos \theta  _1} 
\frac{1}{1/\cos \theta _0 + 1/\cos \theta _1}
\left[ 1 - \exp \left\{-\tau \left( \frac{1}{\cos \theta _0} 
+ \frac{1}{\cos \theta _1} \right) \right\} \right],
\end{eqnarray}
or
\begin{equation}
\label{eq:fatm}
f_{{\rm atm}}(\theta _0, \theta _1, \phi ;\lambda ) 
= \frac{\omega \Psi (\Theta )}{\cos \theta _0 
+ \cos \theta_1}\left[ 1 - \exp \left\{-\tau(\lambda ) 
\left( \frac{1}{\cos \theta _0} 
+ \frac{1}{\cos \theta _1} \right) \right\} \right].
\end{equation}
In the above expressions, $\omega $ is a single scattering albedo that
is the ratio of scattering efficiency to total light attenuation
(both scattering and absorption). In what follows, we ignore the effect of
absorption and assume $\omega = 1$.

 The phase function for 
Rayleigh scattering, $\Psi (\Theta )$, is given by
\begin{equation}
\Psi (\Theta ) = \frac{3}{16\pi }(1 + \cos ^2 \Theta ),
\end{equation}
where $\Theta $ is the angle between the incident and scattered
directions:
\begin{equation}
\cos \Theta = \cos \theta _0 \cos \theta _1 
+ \sin \theta _0 \sin \theta _1 \cos \phi.
\end{equation}
The optical depth for Rayleigh scattering, $\tau $, is empirically given
for the atmosphere of the Earth \citep[e.g., ][]{frohlich1980,young1980}:
\begin{eqnarray}
\tau  &=&  0.00864 \left( \frac{P}{1013.25\;{\rm mbar}} \right)
\tilde \lambda ^{-(3.916 + 0.074\tilde \lambda + 0.05/\tilde \lambda )} \propto \tilde \lambda ^{-4}, 
\label{eq:tau}
\end{eqnarray}
where $\tilde \lambda $ is wavelength in units of 1 $\mu {\rm m}$, 
and $P$ is the pressure of the atmosphere at the base and we set 
$P = 1013.25\;{\rm mbar}$ everywhere for simplicity. The fact that $\tau $ 
is much less than unity in the visible and the near-infrared bands 
justifies the single-scattering approximation, that is adopted in this paper.

\subsubsection{$f_{\rm surf}$ ---Scattering by Surface}

The radiance of the surface, $L_{{\rm surf}}$, is affected by 
atmospheric scattering as well, and is calculated by
\begin{equation}
L_{\rm surf}(\theta _0, \theta _1, \phi ;\lambda ) 
= F_*(\lambda ) \cos \theta _0 \; 
f_{\rm surf}(\theta _0, \theta _1, \phi ;\lambda ), 
\end{equation}
where
\begin{equation}
f_{\rm surf}(\theta _0, \theta _1, \phi ;\lambda ) 
= C_{{\rm atm}}(\theta _0, \theta _1, \phi ; \lambda ) 
f_{\rm surf,\;0}(\theta _0, \theta _1, \phi ; \lambda ),
\end{equation}
$f_{\rm surf,\;0}$ is the BRDF of the surface itself, and $C_{\rm atm}$ 
describes attenuation due to atmospheric scattering before and after 
the ray reaches the surface:
\begin{equation}
\label{eq:C}
C_{\rm atm} = \exp \left\{ -\tau (\lambda ) 
\left( \frac{1}{\cos \theta _0} + \frac{1}{\cos \theta _1} \right) \right\}. 
\end{equation}
We consider the contributions of land, $f_{\rm land,\;0}$, and
ocean, $f_{\rm ocean,\;0}$, in Section \ref{subsubsec:landBRDF} and 
Section \ref{subsubsec:oceanBRDF}, respectively.

\subsubsection{$f_{\rm land,\;0}$ --- Scattering by Land}
\label{subsubsec:landBRDF}

Previous authors \citep{ford2001, tinetti2006a, tinetti2006b,
rodriguez2006, palle2008, oakley2009} first classified the pixels 
on the surface into different surface types, and then computed 
the scattering due to each pixel according to its surface type. 
We do not apply any such classification, but rather use a 
parametrized BRDF for each pixel. More specifically, we adopt 
the model of MODerate resolution Imaging Spectroradiometer 
(MODIS; \citet{salomonson1989}), an instrument on board the EOS/TERRA and
AQUA satellites.  The MODIS team processed the data using the Rossi-Li
model for the BRDF, and assigned wavelength-dependent coefficients
in this model to each observed pixel.  The Rossi-Li model 
is one of semi-empirical kernel-based BRDF models and has been applied widely
\citep[e.g.,][]{lucht2000}. It consists of three terms:
\begin{equation}
\label{eq:Rossi-Li}
f_{{\rm land(RL)}} (\theta _0 , \theta _1 , \phi ;\lambda )
= f_{{\rm iso}}(\lambda ) 
+ f_{{\rm vol}} (\lambda ) K_{{\rm vol}}(\theta _0, \theta _1 , \phi )
+ f_{{\rm geo}} (\lambda ) K_{{\rm geo}}(\theta _0 , \theta _1 , \phi ).
\end{equation}
The first term represents the isotropic component. The second term (the
volume-scattering term) represents the effect of the finite thickness of
the scattering body. The last term (the geometric-optical term) takes
into account of the effect of shadow.  The explicit expressions for $K_{\rm
vol}$ and $K_{\rm geo}$ are given in Appendix \ref{ap:landBRDF}.

In this paper, we adopt the three coefficients in the Rossi-Li model
($f_{\rm iso}, f_{\rm vol}, f_{\rm geo}$) on each $2.5 ^{\circ}\times
2.5 ^{\circ}$ pixel from the dataset ``snow-free
gap-filled MODIS BRDF Model Parameters.''  
This dataset is a spatially and
temporally averaged product derived from the 0$^\circ$.05 resolution
BRDF/albedo data (v004 MCD43C1)\footnote{The data-set is available
through the MODIS web page http://modis.gsfc.nasa.gov/}. 
We select the BRDF parameters evaluated at the central position of each 
$2.5 ^{\circ}\times 2.5 ^{\circ}$ pixel and employ the values to represent 
the pixel. 

The upper three panels of Figure \ref{fig:contour} show the different
behaviors of the surface brightness for each term in the
Rossi-Li model. The lower-left panel plots $f_{{\rm land(RL)}} (\theta
_0 , \theta _1 , \phi ;\lambda )$ for $f_{\rm iso}=0.236$, $f_{\rm
vol}=0.114$ and $f_{\rm geo}=0.027$ (these values are obtained by 
averaging over the pixels corresponding to land).  Note that $f_{\rm land(RL)}$ 
computed from Equation (\ref{eq:Rossi-Li}) is not positive definite, and 
becomes negative for very small $\theta _0 $ or $\theta _1$.  Thus, we set 
$f_{\rm land(RL)}$ to 0 wherever it becomes negative.  The lower-right 
panel of Figure \ref{fig:contour} shows the scattering surface 
brightness due to the ocean described below.

\subsubsection{$f_{\rm ocean,\;0}$ --- Scattering by Ocean
\label{subsubsec:oceanBRDF}}

The data set ``snow-free gap-filled MODIS BRDF Model Parameters''
does not assign parameters ($f_{\rm iso}, f_{\rm vol}, f_{\rm geo}$) for
ocean pixels nor for pixels around the polar region. Our current model 
regards those pixels as ocean. Thus, our model systematically 
underestimates the snow and/or ice areas around the poles. 
Those components actually vary significantly from season to season as well, 
which is not yet taken into account in our present model either. 

Since the scattering pattern by a liquid is very different from that by
land, the above Rossi-Li model (Equation (\ref{eq:Rossi-Li})) is not 
relevant for oceans. Instead, we adopt the ocean BRDF model of 
\citet{nakajima1983}. We describe their model below. 
Further details are discussed in Appendix \ref{ap:oceanBRDF}.

Their model approximately computes the scattering of ocean by the sum of
contributions from small facets (Appendix \ref{ap:oceanBRDF} and
Figure \ref{fig:config_ocean}).  The facets are characterized by the angle of their 
normal directions $\theta_n$, and the scattering for each facet
follows the simple Snell-Fresnel law. The distribution function
$p(\theta_n)$ for the slope of those facets depends on the wind speed above the ocean
(Equations (\ref{eq:thetan-pdf}) and (\ref{eq:oceansigma})).  Thus, their BRDF
model is expressed as
\begin{equation}
\label{eq:oceanBRDF}
f_{\rm ocean}(\theta _0,\theta _1 ,\phi ; \lambda ) 
= \frac{1}{4 \cos \theta _0 \cos \theta _1 \cos \theta _n ^*} 
p(\theta _n ^*) 
G(\theta _0, \theta _1, \phi, u_{10}) 
r(\theta _0 , \theta _1, \theta _n^*, \phi, \tilde m),
\end{equation}
where $\theta _n^*$ indicates the direction of the wave slope
responsible for the specular reflection, $u_{10}$ is the wind speed at
10 m height above the surface, and $\tilde m = \tilde m(\lambda )$ is the
ratio of the refractive index of ocean to atmosphere.  The term
$r(\theta _0 , \theta _1, \phi , \tilde m ) $ stands for Fresnel's
scattering coefficient, $p(\theta _n )$ is the density distribution
function of the wave slope, and $G( \theta _0,\theta _1, \phi, u_{10})$
represents the bidirectional shadowing effect.

Our simulation adopts $u_{10}=4\;{\rm m\,s}^{-1}$, which corresponds to
the spatial average of the wind speed at 10 m above the ocean, 
which is the average of monthly wind speed\footnote[2]{the NOAA-ESRL Physical
Sciences Division, Boulder Colorado. http://www.cdc.noaa.gov/}.
The color of the ocean is affected by the scattering of light inside the
ocean, but we neglect this process for simplicity. We plan to
incorporate it in future work.

\begin{figure}[!h]
  \centerline{\includegraphics[width=12.0cm]{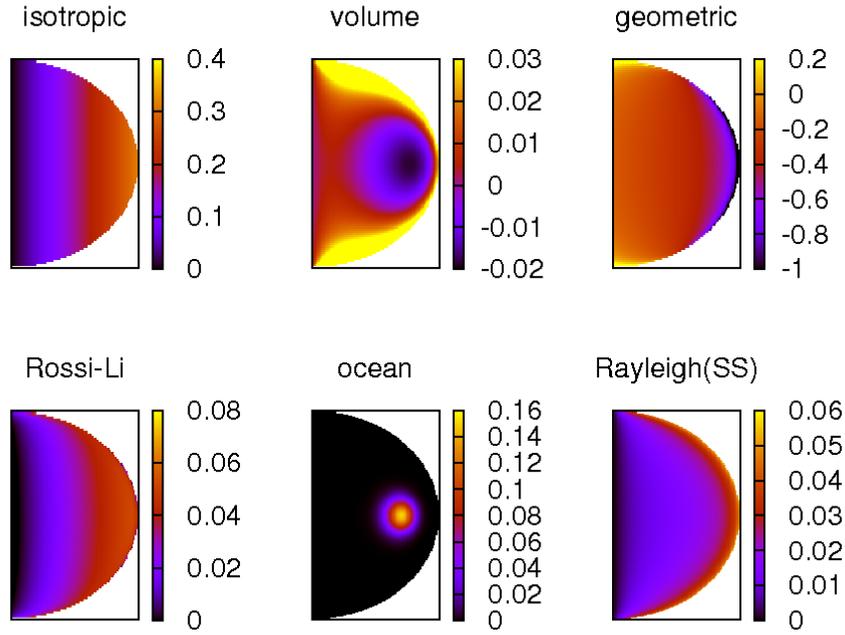}}
\caption{
Surface brightness of the sunlit hemisphere of the planet seen at 
$\alpha = \pi /2$ normalized by solar flux. 
The incident ray is coming from the right. 
The ``isotropic'' panel shows the surface
brightness due to the isotropic term (assuming $f_{\rm iso} =
1$), the ``volume'' panel is due to the $f_{\rm vol}K_{\rm vol}$ term (assuming
$f_{\rm vol}=1$), the ``geometric'' panel is due to the $f_{\rm geo}K_{\rm
geo}$ term (assuming $f_{\rm geo}=1$), and the ``Rossi-Li'' panel is due to
the combination of the three terms, $f_{\rm RL}$, assuming ($f_{\rm iso},
f_{\rm vol}, f_{\rm geo}$)=(0.236, 0.114, 0.027), a set of averaged values 
over the entire land surface.  The ``ocean'' panel is based on the ocean BRDF 
model by \citet{nakajima1983} with a wind velocity $4\;{\rm m\,s}^{-1}$.
The ``Rayleigh (SS)'' panel is the surface brightness due to Rayleigh scattering 
of the atmosphere within a single-scattering (SS) and a flat atmospheric 
layer approximations, which are quantitatively not accurate at the edges, 
where either $\theta _0$ and $\theta _1$ are close to zero.
} \label{fig:contour}
\end{figure}

The surface brightness of ocean in this model is shown in the lower
middle panel of Figure \ref{fig:contour}.  The color contours clearly exhibit  
the remarkable increase in reflectance of the ocean near $\theta _0 = \theta _1$ due to 
specular reflection.  The strong specular reflection feature will be further
discussed in Section \ref{s:tau}.

\begin{deluxetable}{lcc}
  \tablewidth{0pt}
  \tablecolumns{3}
 \tablecaption{Canonical Values of Parameters Assumed in Our Simulations
of Photometric Light Curves.
  \label{tab:param}}
  \tablehead{
    \colhead{Parameter}& \colhead{Symbol} & \colhead{Value}
  }
  \startdata
  Planet-observer distance & $l$ & 10 pc \\
  Effective diameter of telescope & $D$ & 2 m \\
  Exposure time & $t_{\rm exp}$ & 1 hr \\
  Observation period & $n$ & 14 days \\
  \enddata
\end{deluxetable}

Adopting the above scattering model, we are now able to compute
scattered light curves of a ``Second Earth'' (but without clouds).  
The planet rotates around its spin axis with a period of 24 hr. The
parameters of our mock observations are listed in Table \ref{tab:param}.

We ignore the effect of the spin rotation during the exposure time
$t_{\rm exp}$. The orbital motion of the planet is completely ignored
during the observation period $n$.  We employ the set of MODIS
photometric bands on which our data for the land BRDF are based.  These bands
are 0.459--0.479 $\mu $m (band 1), 0.545--0.565 $\mu $m (band 2),
0.620--0.670 $\mu $m (band 3), 0.841--0.876 $\mu $m (band 4),
1.230--1.250 $\mu $m (band 5), 1.628--1.652 $\mu $m (band 6), and
2.105--2.155 $\mu $m (band 7).  In reality, we do not integrate over
the band but simply calculate the scattering at the central wavelength of
each band and multiply it by the band width.

In practice, the spin rotation period of the planet is unknown a priori,
and we must first determine it from the observed light
curves. \citet{palle2008} discussed how to infer the spin rotation
period from the photometric variation and found that auto-correlation
analysis of light curves can effectively determine the period. In this
paper, therefore, we assume that the spin rotation period of the planet
is precisely known, and fold the light curves accordingly.

\subsection{Results \label{s:result}}
\begin{figure}[!tbh]
  \centerline{\includegraphics[width=16.0cm]{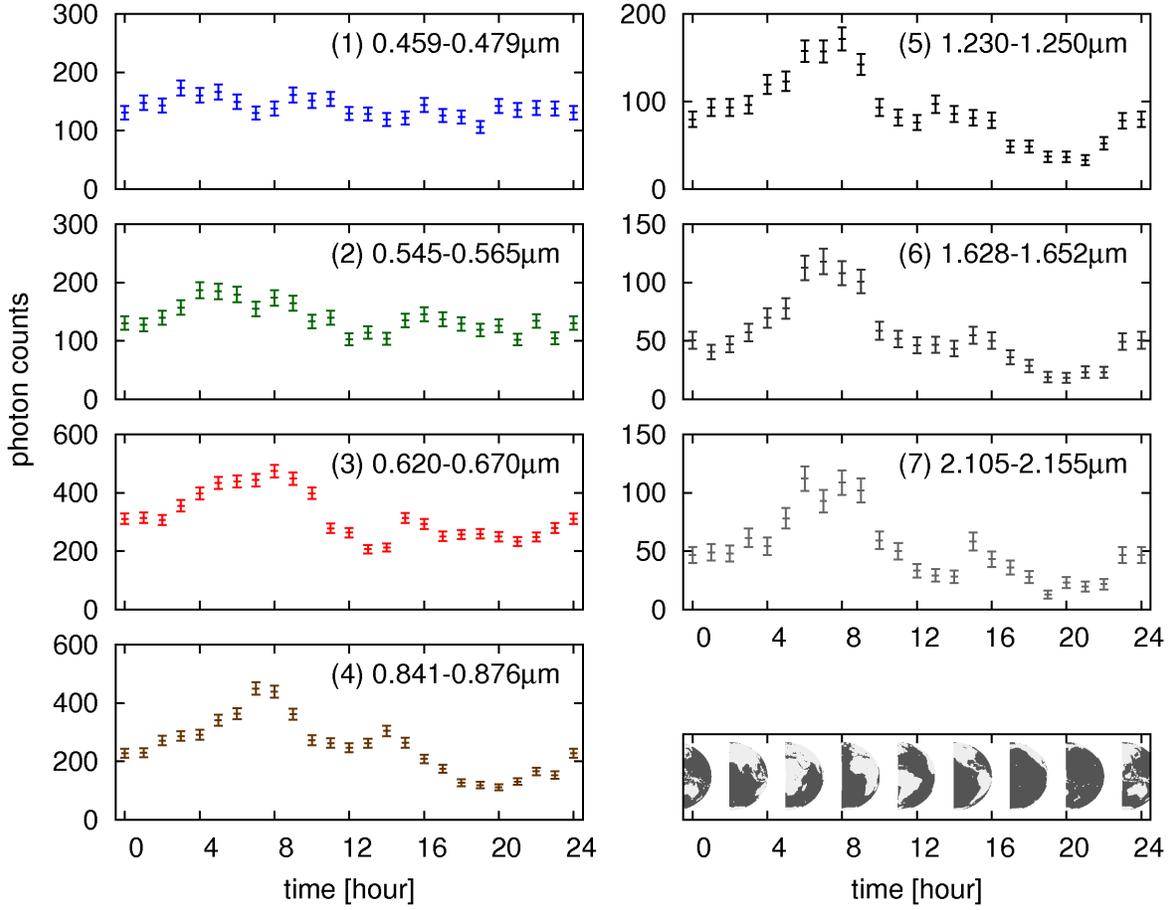}}
\caption{Light curves of the cloudless Earth at 0.459--0.479 $\mu $m
(band 1), 0.545--0.565 $\mu $m (band 2), 0.620--0.670 $\mu $m (band 3),
0.841--0.876 $\mu $m (band 4), 1.230--1.250 $\mu $m (band 5),
1.628--1.652 $\mu $m (band 6) and 2.105--2.155 $\mu $m (band 7). The error
bars come from photon shot noise only.  The bottom right panel shows the
snapshots of the Earth at corresponding epochs. The ocean is painted in
black and the land is in white.  The observer is located on celestial
equator and half of the projected planetary surface is illuminated.  }
\label{fig:scatter2}
\end{figure}

Figure \ref{fig:scatter2} shows the simulated light curves in the seven 
bands. We consider a very idealized observational situation in which the light from the
host star is completely blocked, and the photometric noise is due to the Poisson
fluctuations in the observed photon counts from the planet alone:
\begin{equation}
N \sim 840 \left( \frac{I}{10^{15}\;{\rm W\,str^{-1}}\mu{\rm m}^{-1}} \right) 
\left( \frac{l}{10\;{\rm pc}} \right)^{-2} 
\left( \frac{D}{2\;{\rm m}}\right)^2 
\left( \frac{t_{\rm exp}}{1\;{\rm hr}} \right) 
\left( \frac{n}{14\;{\rm days}} \right)  
\left( \frac{\lambda }{1\;\mu {\rm m}} \right) 
\left( \frac{\Delta \lambda }{0.1\;\mu{\rm m}} \right).
\end{equation}
Here, the errors scale as $\sim \sqrt{\rm N}$.  In reality, however, there are
certainly many sources of errors such as contamination by the host star, zodiacal light 
and detector noise; furthermore the flux from the planet will be 
attenuated through the telescope optical instrument and detectors.
However for this
preliminary investigation, we consider this idealized observational situation 
and leave a more realistic
treatment of errors to future studies.

Light curves at short wavelengths (especially in bands 1 and 2) in Figure 
\ref{fig:scatter2} do not exhibit significant time variation because 
Rayleigh scattering by the atmosphere is dominant at shorter wavelengths and 
dilutes the variations of surface features.  The atmosphere becomes significantly 
more transparent at longer wavelengths and the variations of light curves due to the
inhomogeneous surface become appreciable.  A comparison of the
light curves with the snapshots in Figure \ref{fig:scatter2} indicates
that the three bumps at $t \sim 1, 8,$ and $14$ hr indeed correspond
to the Eurasian, African, and American continents. The highest peak shows up
when the Sahara desert emerges.  A dip at $t \sim 20$ hr occurs
when the illuminated and visible part is covered with ocean, since the
reflectivity of land is higher than that of ocean, especially at longer
wavelengths.  Another dip at $t \sim 13$ hr occurs when the South
American continent replaces the specular reflection point.  All these
features are consistent with the light curve simulation of a cloudless
Earth by \citet{ford2001} and \citet{oakley2009}, and are essential in 
extracting information concerning the surface features of a planet from its 
light curves.

\begin{figure}[!tbh]
  \centerline{\includegraphics[width=14.0cm]{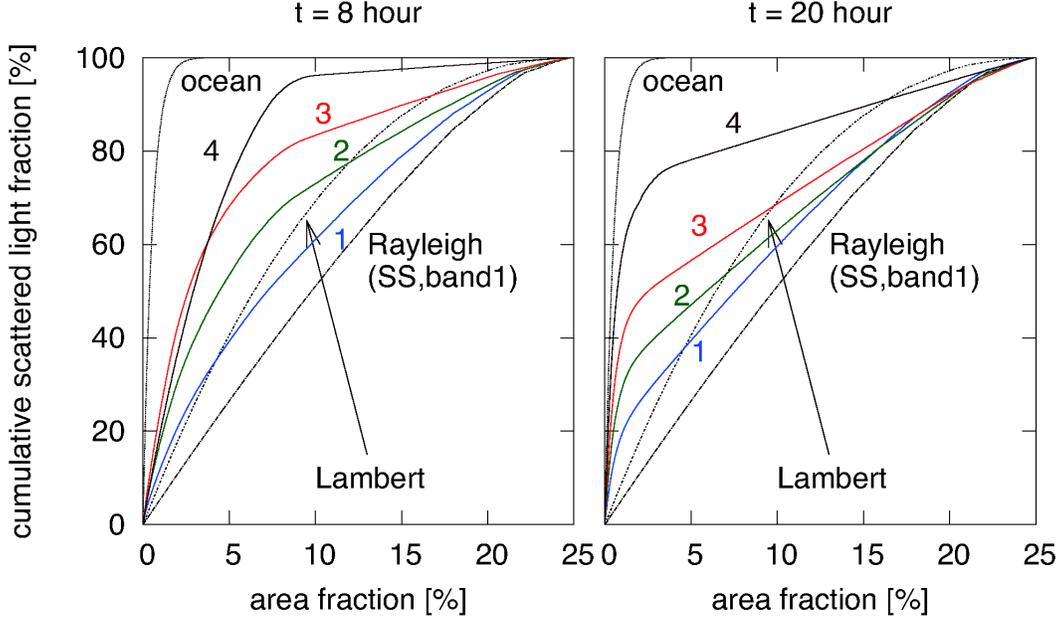}}
\caption{Scattering properties of different types of surfaces. Cumulative
fraction of scattered light is plotted against the corresponding fractional 
area defined as responsible for the scattered light.  The lines at $t =
$8, 20 hr are plotted with three references ---ocean with wind velocity 
$u_{10} = 4\;{\rm m\,s}^{-1}$, Rayleigh scattering and Lambertian.  
Different colors represent different bands as in Figure \ref{fig:scatter2} 
(band 1: blue, band 2: green, band 3: red, band 4: brown).  The light 
scattered by the ocean is very localized.}  
\label{fig:conc}
\end{figure}

Figure \ref{fig:conc} illustrates the degree of localization of the
source of the observed flux 
from the planetary surface. According to the
geometry of the system that we adopt here, a quarter of the planetary
surface area is illuminated and visible to the observer. We sort all the pixels located
in the illuminated and visible part of the surface according to the amount 
of scattered light per area, and compute the cumulative factional scattered 
light as a function of the corresponding fractional area. 
In Figure \ref{fig:conc}, if all the pixels contribute to the
light equally, the resulting plot would be a straight line 
reaching 100\% at the fractional area of 25\%. In
reality, however, the plot is slightly curved even in the case of
isotropic scattering (Lambertian) due to the curvature of the
global surface (Figure \ref{fig:contour}). 
Since light from the ocean region   
comes from a very localized specular spot, the dotted line in Figure 
\ref{fig:conc} is very steep at small fractional area ---the ocean spot equivalent
to $\sim $ 2\% of the surface area is responsible for nearly the entire 
light from the ocean.

\begin{figure}[!tbh]
  \centerline{\includegraphics[width=10.0cm]{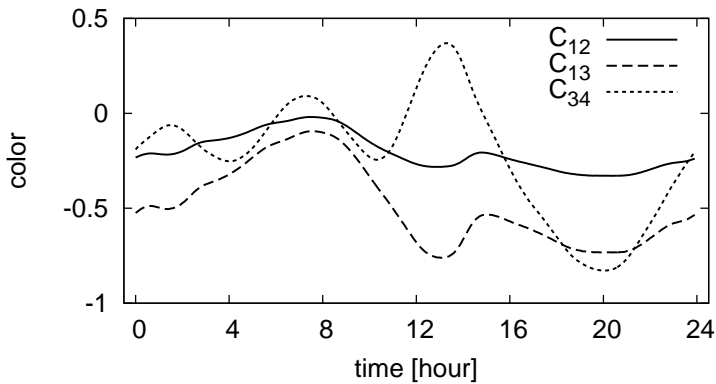}}
\caption{Variation of colors of Earth against time. 
 Considered bands are 0.459--0.479 $\mu $m (band 1), 0.545--0.565 $\mu $m (band 2), 
 0.620--0.670 $\mu $m (band 3), and 0.841--0.876 $\mu $m (band 4). Here, ``color'' 
 of band $a$ and $b$ is defined as $C_{ab} \equiv -2.5 \log \frac{I_a}{I_b}$ where 
 $I_a$ and $I_b$ are the energy fluxes per wavelength in band $a$ and band $b$. 
}
  \label{fig:color}
\end{figure}

\begin{figure}[!tbh]
  \centerline{\includegraphics[width=8.0cm]{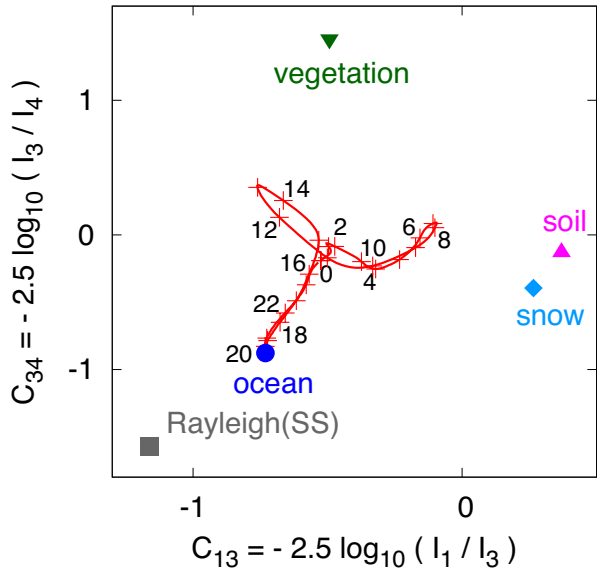}}
\caption{Noiseless light curve trajectory on a color-color diagram assuming a 
black-body spectrum as the incident flux. The ``color'' of band $a$ and $b$ is 
defined as $C_{ab} \equiv -2.5 \log \frac{I_a}{I_b}$ where $I_a$ and $I_b$ 
are the energy fluxes per wavelength in band $a$ and band $b$. 
The numbers denote the time (hours) corresponding to those in Figure \ref{fig:scatter2}. 
The symbols labeled by ``ocean,'' ``soil,'' ``vegetation,'' ``snow,'' and 
``Rayleigh (SS)'' indicate the locations of the typical scattering colors of 
that surface type with atmospheric layer on it. Wavelength-dependent albedo of 
these five types are exhibited in Figure \ref{fig:effalbedo}.}  
\label{fig:colorcolor}
\end{figure}

Essentially, the information that we utilize in reconstructing the fractional 
areas is color variations at different phases of the surface. 
To put it more clearly, we define the {\it color} between band $a$ and band $b$ as
\begin{eqnarray}
C_{ab} \equiv -2.5 \log \frac{I_a}{I_b},
\end{eqnarray}
where $I_a$ and $I_b$ are 
the energy fluxes per wavelength in band $a$ and band $b$ (Equation (\ref{eq:intensity})).
Figure \ref{fig:color} shows time variations of $C_{12}$, $C_{13}$, and
$C_{34}$. The trajectory on the $C_{13} - C_{34}$ plane is plotted in Figure
\ref{fig:colorcolor} together with the typical colors of ocean, snow, vegetation, 
soil, and Rayleigh scattering which will be described in the next section and 
in Figure \ref{fig:effalbedo}. The trajectory indicates that the illuminated 
and visible part of the surface is almost fully covered by ocean at $t=20$ hr. 
The other tips at $t=8$ and $13$ hr correspond to the African continent with 
the Sahara desert and the South American continent with the Amazon forest, respectively.
These color variations 
play a key role in estimating fractional areas of different surface types.

\section{Reconstruction of the fractional areas of surface types}
\label{s:inverse}

In this section, we describe our methodology for reconstructing the
fractional areas of different surface types from multi-band photometry and
present the results from the analysis of the mock light curves described
in Section \ref{s:simulation}.

\subsection{Inversion Method \label{s:method}}
\label{ss:inversion}

Our basic strategy for reconstructing the surface features of a planet
is to fit the photometric light curves with an {\it a priori} model of planetary
scattering. In doing so, we adopt two major simplifying assumptions.
One is the Lambertian model  (i.e., isotropic scattering)  for the
surface, and the other is that the surface consists solely of four different types
(ocean, soil, vegetation, and snow) plus an atmosphere.

The Lambertian surface is one of the simplest models of scattering with
constant radiance against any geometry of incident and scattered rays.
The BRDF for a Lambertian surface is simply given as
\begin{equation}
f_k(\theta _0, \theta _1, \phi ;\lambda ) 
= f_{{\rm iso}\;k}(\lambda ) = \frac{a_k(\lambda )}{\pi }, 
\end{equation}
where the subscript $k$ denotes an index of the surface types and
$a_k(\lambda )$ is a wavelength-dependent albedo of the $k$-th surface type.

The validity and limitation of our reconstruction method crucially
depend on the number of different surface types that we consider.  In
practice, however, the limited information of color variations strongly
restricts the number that can be uniquely determined by the analysis.  Thus we
consider only four types ($k=$ ocean, soil, vegetation, and snow)
that constitute the major components of the surface of the Earth.

\begin{figure}[!tbh]
  \centerline{\includegraphics[width=9cm]{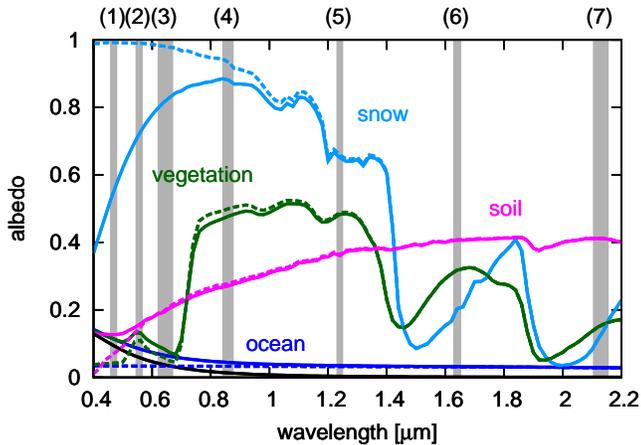}}
\caption{Wavelength-dependent effective albedos of ocean (blue), soil
(magenta), vegetation (green), snow (cyan), and atmosphere with Rayleigh
scattering alone (black).  The solid lines show the effective albedo
with Earth-like atmosphere, while the dashed lines show the effective
albedo without an atmosphere. 
Shaded regions correspond to the MODIS bands. The numbers at the top are 
the labels of the different photometric bands.}  
\label{fig:effalbedo}
\end{figure}

This approach inevitably limits the generality of our model.  Nevertheless
we think it reasonable for the present purposes for several reasons. 
First, our Earth is currently the only known example of a habitable
planet, and it is not unreasonable to assume that at least some habitable
exoplanets have similar surface properties. 
Second, these four types represent very different albedos 
(Figure \ref{fig:effalbedo}), which makes it easier to
distinguish them in scattered light.  Finally, the presence of ocean(s)
on a planet is
deeply related to the fundamental question of its habitability and the 
vegetation red edge can be regarded, if it exists at all, as a
direct indication of the presence of life.

Under the conditions described above, Equation (\ref{eq:intensity}) at a
given epoch $t_i$ reduces to
\begin{eqnarray}
I(\lambda ) 
&=&  F_* (\lambda ) R_{\rm p} ^2 
\int _S f_{{\rm iso}\;k}(\lambda ) \cos \theta _0 \cos \theta _1 ds 
\nonumber \\
&=& F_* (\lambda )  R_{\rm p} ^2 
\sum _{k}\left\{ \frac{a_k(\lambda )}{\pi }  
\int _{s_k} \cos \theta _0 \cos \theta _1 ds \right\},
\label{eq:ilambda}
\end{eqnarray}
where the integration is performed over $s_k=s_k(t_i)$ that is the area of the
$k$-th surface type in the illuminated and visible area at $t_i$.
Denoting each band by an index $j$ ($=1,2...,7$), we discretize Equation
(\ref{eq:ilambda}) as
\begin{eqnarray}
I_{j}(t_i) &=& F_{*j} R_{\rm p} ^2  \int _s \cos \theta _0 \cos \theta _1 ds \;  
\sum _{k} D_{jk}A_{k}(t_i), \label{eq:inv}\\
D_{jk} &\equiv & f_{{\rm iso}\;\;jk} = \frac{a_{jk}}{\pi},\label{eq:D}
\\
A_{k}(t_i) &\equiv & \frac{\displaystyle \int _{s_k(t_i)} \cos \theta _0 \cos \theta _1 ds}
{\displaystyle \int _s \cos \theta _0 \cos \theta _1 ds},
\label{eq:A}
\end{eqnarray}
where $A_k$ is the geometrically-weighted fractional area of the $k$-th 
surface type that we want to estimate, and $D_{jk}$ is normally referred 
to as the ``design matrix" by statisticians.

In Equation (\ref{eq:inv}), the incident flux $F_{*j}$ is calculated
from the intensity of the host star and the distance between
the star and the planet $d$, which is obtained 
once the orbit of the planet is determined. 
The integral $\int _s \cos \theta _0 \cos
\theta _1 ds $ depends on the phase angle $\alpha $ alone (in
our current configuration of the phase angle $\alpha = \pi /2 $, 
$\int _s \cos \theta _0 \cos \theta _1 ds = 2/3$).

In order to estimate the weighted fractional area $A_k(t_i)$, we need to 
determine the design matrix, $D_{jk}$ from the albedo $a_{jk}$.  We define
the effective albedo of the $k$-th surface type, $a_{{\rm eff}\;k} (\lambda)$, 
by setting it equal to the actual reflectivity when the whole
surface is covered by that surface type.  More specifically, it is
computed as
\begin{equation}
a_{{\rm eff}\;k}(\lambda ) \equiv \pi \;  
\frac{\displaystyle 
\int _s f_{{\rm model}\;k}(\theta _0 , \theta _1, \phi ;\lambda ) 
\cos \theta _0 \cos \theta _1 ds}
{\displaystyle \int _s \cos \theta _0 \cos \theta _1 ds} ,
\label{eq:effalbedo}
\end{equation}
where $f_{{\rm model}\;k}$ is the model BRDF of the $k$-th surface type 
and the integrations are performed over the illuminated and visible area
which, again, depends on the phase angle $\alpha $ alone. Then we define
$a_{jk}$ of the $j$th band from its central wavelength, $\lambda_j$, as
\begin{equation}
a_{jk} = a_{{\rm eff}\;k}(\lambda_j) .
\end{equation}
The model BRDF $f_{{\rm model}\;k}$ in Equation (\ref{eq:effalbedo}) is
calculated from
\begin{equation}
f_{{\rm model}\;k}(\theta _0, \theta _1, \phi ;\lambda ) 
=  f_{{\rm atm}}(\theta _0, \theta _1, \phi ;\lambda ) 
+ C_{{\rm atm}}(\theta _0, \theta _1, \phi ;\lambda ) 
f_{{\rm surf,0}\;k}(\theta _0, \theta _1, \phi ;\lambda ),
\label{eq:fmodel}
\end{equation}
where $f_{\rm atm}$ and $C_{{\rm atm}}$ are given by Equations
(\ref{eq:fatm}) and (\ref{eq:C}), respectively. 
In order to determine $f_{{\rm surf,0}\;k}$ of the three land components 
($k=$ soil, vegetation and snow), we first assume that they are Lambertian 
with scattering spectra given by the ASTER spectral library
\footnote[3]{http://speclib.jpl.nasa.gov/} \citep{aster}. 
Specifically, we adopt ``Brown to dark brown sand (Entisol),'' 
``Grass,'' and ``Fine Snow'' for soil, vegetation, and snow, respectively.  
Since the ASTER spectral library offers data at discrete wavelengths, 
we linearly interpolate the data to obtain the $f_{{\rm surf,0}\;k}$ 
suitable for our MODIS bands. For $f_{{\rm surf,0}\;k}$ of ocean, we
use Equation (\ref{eq:oceanBRDF}).  Figure \ref{fig:effalbedo} displays 
the effective albedo at $P=1013.25$ mbar for ocean, soil, vegetation, 
and snow as a function of wavelength $\lambda $.  The dashed lines are the 
albedo curves in the absence of an atmosphere.  The black solid line is 
the effective albedo due to atmospheric Rayleigh scattering. 
The primary effect of Rayleigh scattering is to add a very blue continuum 
to every pixel's contribution to the total light, but in our single-scattering 
approximation it also reduces the amount of light reaching the surface. 
In Figure \ref{fig:effalbedo}, this is most clearly visible in the 
blue spectral region for the snow component. 

\subsection{Fitting}
\label{s:fitting}

Now we apply the above methodology to the mock light curves
(Section \ref{s:simulation}).  We determine the best-fit values of $A_j (t_i)
$ in Equation (\ref{eq:inv}) by chi-square fitting:
\begin{equation}
\label{eq:chi}
\chi ^2(t_i) = \sum _j \frac{\left\{ I_{{\rm obs}\;j} (t_i) 
- F_{*j} R_{\rm p} ^2 \int _s \cos \theta _0 \cos \theta _1 ds \; 
\sum _{k} D_{jk}(\tau ) A_{k}(t_i)\right\} ^2 }{\sigma ^2 _{j}(t_i)}.
\end{equation}
The fit is independently performed for the light curves at each epoch
$t_i$.  In order to quantify the errors of fitted values, we generate
100 realizations by adding a Poissonian error with rms of
$\sigma_j$ to the simulated data, and compute the average and the
variance of the best-fit values.  We consider the photon shot noise
alone in $\sigma _j$, ignoring the other statistical errors
of the phase angle $\alpha $, star-planet distance $d$, and
planet-observer distance $l$ among others. Thus our model is admittedly
very idealized but indicates what one can learn from possible future data in
principle.

Equation (\ref{eq:chi}) is essentially a linear inverse problem. So as
to ensure the positivity in each element of $A_j$, however, we replace
$A_j$ by $B_j^2$ and search for the best-fit $B_j$ with a nonlinear
fitting method \citep[the Levenberg-Marquardt method, e.g.,][]{nrecipe}. 
We also made sure that another independent fitting method, which is 
based on a conditional least square method and does not use the above 
replacement \citep[e.g.,][]{menke1989}, gives very similar results.

Our model does not assume the condition $\sum _{k=1}^4 A_k = 1$.  We
could impose that constraint in principle, but it would somewhat restrict
the applicability of the model.
After all, the actual planetary surface does not exactly consist of only
four Lambert surface types as we assumed here, and
components other than the four types might contribute.
In addition, the sum of the estimated
area can easily deviate from unity due to a variety of other
effects including the anisotropy of the scattering and the diversity of
the detail features of the reflection spectra (e.g., the sharpness
of the red edge, the slope of the soil spectrum etc.),
Therefore, we fit the mock data without any constraints other than
$A_k(t_i) > 0$.

\subsection{Estimation of the Weighted Fractional Areas}
\label{s:results2}

In reality, the atmospheric optical depth due to Rayleigh scattering $\tau $ will not
be known in advance and should be fitted simultaneously from the data.  In
this subsection, however, we simply set the optical depth to the
standard value $\tau _0$:
\begin{equation}
\tau _0 (\lambda ) = 0.00864 \tilde \lambda ^{-4},
\label{eq:tau0}
\end{equation}
which is an approximation form of Equation (\ref{eq:tau}) with $P=$1013.25
mbar.  The effect of $\tau $ will be discussed in Section \ref{s:tau}.

\begin{figure}[!htbp]
\centerline{\includegraphics[width=12.0cm]
{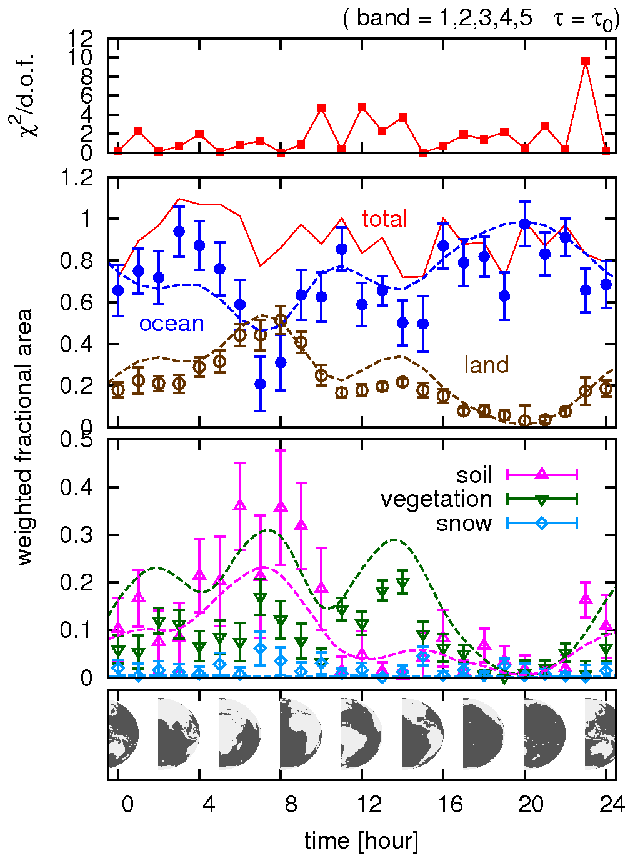}} 
\caption{Reconstructed fractional areas $A_k(t_i)$ for four surface types 
from the simulated light curves in five bands (bands 1 - 5). 
The top panel shows the value of reduced $\chi ^2$ (=$\chi ^2/{\rm dof}$ 
where degree of freedom (dof) is $5-4=1$) for each epoch. 
The upper middle panel displays the results of estimating weighted fractional 
areas of ocean (blue), land (=soil+vegetation+snow; brown), and the total of them (red). 
The lower middle panel displays those of soil (magenta), vegetation (green), 
and snow (cyan).  The dashed lines
in those two panels show the weighted fractional areas derived from the 
MODIS land ecosystem classification dataset. 
The quoted error bars indicate the variance of the best-fit values from 
100 realizations. The bottom panel depicts the snapshots of the Earth
at the corresponding epochs where the ocean is painted in gray and the
land in white.}  \label{fig:inv_scatter2}
\end{figure}

The top panel of Figure \ref{fig:inv_scatter2} shows the $\chi^2/{\rm d.o.f.}$ 
$({\rm d.o.f.}=5-4=1)$ of the fitting at each epoch. 
The second panel from the top presents the result of estimating weighted 
fractional areas $A_k(t_i)$ of ocean and land (= soil + vegetation
+ snow) using the light curves in bands 1, 2, 3, 4, and 5. 
The third panel from the top panel displays the fractional areas of the 
three land types separately. The bottom panel illustrates the corresponding 
snapshot of the Earth toward the observer.  The symbols indicate the average 
of the best-fit values from the 100 realizations with quoted error bars 
representing the rms among them.  For reference, we plot in dashed curves the 
weighted fractional areas based on the One-Minute Land Ecosystem Classification
Product, which is generated from the official MODIS land ecosystem
classification dataset classifying the surface of the Earth into 16
classes.  Among the 16 classes, we regard ``water'' as ocean, ``snow and
ice'' as snow, ``open shrubland'', ``permanent wetlands'', ``urban and built-up'', 
and ``barren or sparsely vegetated'' as soil, and others as vegetation, 
as shown in Table \ref{tb:class}.

The comparison of our best-fit values and dotted lines indicates that the 
present method works fairly well. Given the relatively crude approximation 
of the isotropic scattering and the assumption of only four surface types 
incorporated in the analysis, it is perhaps surprisingly successful. 
The presence of ocean, soil and vegetation is recovered, and the variation of their
weighted fractional areas follows ground truth at least qualitatively.  The bumps of
the ocean at $t\sim 11$ and $20$ hr correspond to the Atlantic
Ocean and the Pacific Ocean, respectively.  The peaks in the soil and
vegetation curves indeed correspond to the Sahara desert and the Amazon
forest, respectively. Moreover, the weighted fractional area of snow is 
consistent with zero, in agreement with the fact that we have adopted 
the {\it snow-free} land BRDF data in the light curve simulation.  

The quantitative discrepancy between the expected and model fractional 
areas of soil and of vegetation probably comes from the diversity of land 
scattering properties that is actually not well represented by the
four Lambertian types shown in Figure \ref{fig:effalbedo}.  Note that our
assignment of the 16 surface classification into the four types (Table
\ref{tb:class}) is not unique, and therefore the dashed lines should be
regarded as a plausible but not unique reference.

\begin{center}
\begin{table}
 \caption{The International Geosphere-Biosphere Programme(IGBP)
  classification which is generated from the official MODIS land
  ecosystem classification data-set(MOD12Q1). The fourth column is the
  merged classification we adopt for dashed lines in Figure
  \ref{fig:inv_scatter2}. } \label{tb:class}
\begin{tabular}{ccccc}
\hline \hline
No. &IGBP Classification     & Area(\%) &Our Classification & Ocean/Land \\ \hline
  0	&	water	&71.40& ocean & ocean\\ 
  1	&	evergreen needleleaf forest	&1.13&vegetation & land\\ 
  2	&	evergreen broadleaf forest	&2.87&vegetation &land\\ 
  3	&	deciduous needleleaf forest &0.18&vegetation& land\\
  4	&	deciduous broadleaf forest	&0.46&vegetation& land\\ 
  5	&	mixed forest	 &1.34&vegetation & land\\
  6	&	closed shrubland&0.16& vegetation & land\\
  7	&	open shrubland &5.22& soil & land\\
  8	&	woody savannas &2.15&vegetation & land\\
  9	&	savannas &1.99 &vegetation & land\\
  10	&	grasslands &2.65 & vegetation & land\\
  11	&	permanent wetlands &0.06 &soil &land \\
  12	&	croplands	&2.56 & vegetation & land\\
  13	&	urban and built-up	&0.14 &soil & land\\
  14	&	cropland/natural vegetation mosaic&0.60 &vegetation & land\\
  15	&	snow and ice	&3.16 & snow & land\\
  16	&	barren or sparsely vegetated&3.92 & soil & land\\
  17	&	unclassified&0.00 &- &- \\ \hline
\end{tabular}
\end{table}
\end{center}

Given the fact that scattered light from oceans comes almost exclusively 
from a very small region of specular reflection, it is perhaps 
puzzling that our inversion method, based on an assumption of Lambertian
scattering, can estimate the fractional area of ocean
reasonably well. This can be understood as
follows: except for the specular reflection spot, the scattered
light from the ocean surface is negligible (Figure \ref{fig:contour}).
Rayleigh scattering in the atmosphere adds a uniform and very blue continuum 
to the observed flux from every part of the illuminated and visible part of
the planetary surface, including areas of ocean away from the specular reflection point. 
For land areas, the contribution to the scattered light is dominated
by direct scattering from the surface, especially in the redder bands, so the
effect of Rayleigh scattering is merely to produce a slightly bluer
overall color (Fig. \ref{fig:effalbedo}). 
For ocean areas outside the specular reflection spot, however, the contribution 
to the total scattered light comes primarily from the Rayleigh scattering.  
Furthermore, the resulting scattered light by the atmosphere is fairly 
diffuse and comes from the entire illuminated and visible area. 
Thus, in order to account for the observed amount of Rayleigh scattering color 
in the total, the fit must assign a sufficiently large area of dark (very low albedo) 
area to the surface. 
Ocean is the only low albedo component available to the fit among the four 
components and thus is automatically assigned to produce enough Rayleigh 
scattering color without producing too much land scattering color. 
In other words, in our fit (which makes no use of spatial or time-domain 
information), the Rayleigh scattering component without any corresponding 
land component is a proxy for the ocean. The dark ocean surface is detectable 
in the fit to the colors primarily due to the additional Rayleigh 
scattering continuum supplied by the atmosphere above it.

\subsection{Time-integrated Spectra}
\label{s:integrate}

\begin{figure}[!tbh]
  \centerline{\includegraphics[width=9.0cm]{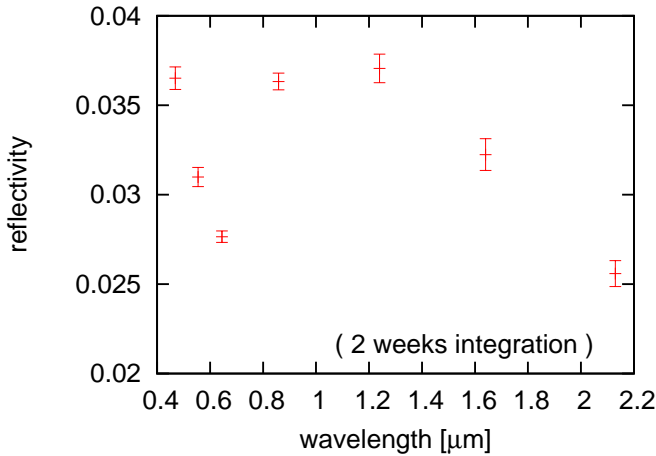}}
\caption{Spectrum of the cloudless Earth from the total integration time (2 weeks). 
The $y$-axis is normalized by the scattered light by a lossless Lambert sphere 
at full phase. }
  \label{fig:sumspec}
\end{figure}

Since our fiducial observational condition (Table \ref{tab:param}) is fairly idealized, 
the quoted errors are small even in the case of time-dependent analysis. 
In an actual missions, however, time-intergrated spectra will probably be the first 
realistic goal.  The spectrum from the total integration time of the mock observation (2 weeks) is displayed in Figure \ref{fig:sumspec}.  
This result is obtained by adding up the photon counts for each band over two
weeks and then normalizing them with respect to the scattered intensity of
a lossless Lambert sphere.  The error bars come from the shot noise of
the total photon count in each band. Note that the red
edge produces a sharp increase at $\lambda \sim 800\,{\rm nm}$.

\begin{figure}[!tbh]
  \centerline{\includegraphics[width=8.0cm]{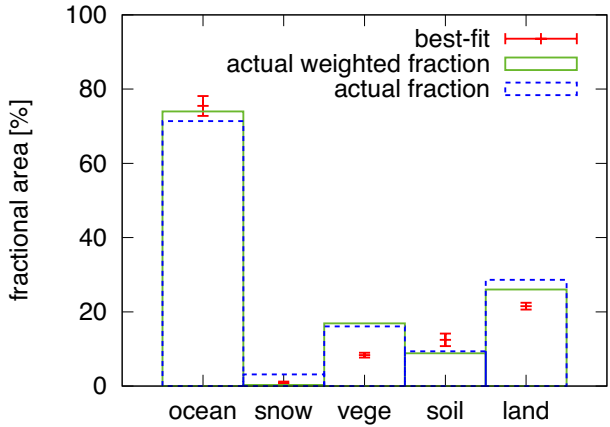}} 
\caption{Time-averaged spectrum from the whole integration time (Figure \ref{fig:sumspec}) 
is decomposed into ocean, snow, vegetation, and soil with Equation (\ref{eq:inv}). 
The land is the summation of snow, vegetation, and soil. The green boxes are the weighted 
fractional areas based on our merged classification (Table \ref{tb:class}) and the blue 
boxes show the non-weighted actual fractions.
}  \label{fig:suminv}
\end{figure}

We present in Figure \ref{fig:suminv} the result of the fit to this
time-averaged spectrum with our fiducial configuration.  The fitting
model is now the time-averaged version of Equation (\ref{eq:inv}) and
thus the estimated values correspond to $\overline{A_k(t_i)}$ averaged over
the full integration time.  The blue boxes represent the actual fractional  areas,
while the green boxes represent the fractional areas based on our merged
classification (Table \ref{tb:class}) but correspondingly
weighted to be compared with the estimated value.  
Under our fiducial configuration, the fractional areas are reasonably reproduced even 
without time resolution (and hence without spatial resolution).

We also repeat the same fitting procedure using mock light curves with
different geometries, and summarize the results in Table
\ref{tab:geometry}.  The values in the ``reference'' lines indicate weighted and 
averaged fractional areas visible for each observer.
The bottom line shows the actual (non-weighted) fraction of the four surface components. 
Of course, the correspondence of the actual non-weighted fractional areas and the 
weighted fractional areas is highly dependent on the geometry. In the case of the Earth, 
the weighted fractions are roughly equivalent to the actual fractions in our fiducial 
configuration i.e. if the Earth is at equinox and the observer is on the 
equatorial plane, due to the fact that areas of ocean, soil, and vegetation are not 
extremely inhomogeneous but distributed well along the latitude. 
On the other hand, land areas would be preferentially seen as viewed by an observer 
located above the northern hemisphere. Although such geometric limitation is inevitable, 
the estimated values recover the weighted fractional areas for each geometry reasonably well.
This is very encouraging for future missions.

\begin{deluxetable}{llccccc}
  \tablewidth{0pt}
  \tablecolumns{7}
 \tablecaption{Estimated values of weighted fractional areas from time-averaged 
 spectra with different geometries. The ``north 45$^{\circ}$'' and ``south 45$^{\circ}$'' 
 assumes the Earth at equinox seen with $+45^{\circ}$ and $-45^{\circ}$ inclination 
 from the equatorial plane, respectively. The ``summer (winter) solstice'' assumes 
 the Earth at summer (winter) solstice and the observer on the intersection of the 
 equatorial plane and orbital plane. The values in ``reference'' lines are based on 
 our merged classification shown in Table \ref{tb:class} which are weighted and 
 averaged according to each geometry. The ``actual (non-weighted)'' is the non-weighted 
 fractional areas of each components based on our merged classification. 
 \label{tab:geometry}}
  \tablehead{
    \colhead{Geometry}	& \colhead{} & \colhead{ocean (\%)} & \colhead{land (\%)} & \colhead{soil (\%)} & \colhead{vegetation (\%)} & \colhead{snow (\%)}
  }
  \startdata
  fiducial &	estimated	& 75.5	&	21.5		& 12.4	&	8.3 &	0.7 \\ \cline{2-7}
  &	reference	& 74.0	&	26.0 &	8.8 &	16.8 &	0.3  \\ \hline  
  north $45{^\circ}$ &	estimated	& 66.5	&	29.4		& 17.5	&	10.1 &	1.7 \\ \cline{2-7}
  &	reference	& 56.9	&	43.1 &	17.2 &	25.1 &	0.8  \\ \hline
  south $45{^\circ}$ &	estimated	& 86.6	&	10.7	 &	4.3 	& 5.6	&	0.7 \\ \cline{2-7}
  &	reference	& 85.6	&	14.3 &	4.0 &	6.9 &	3.4  \\ \hline
  summer solstice &	estimated	& 73.1	&	26.5	 &	16.3 & 9.4	& 0.8	\\ \cline{2-7}
  &	reference	& 68.0	&	32.3 & 11.5	&	20.3	 &	0.03  \\ \hline
  winter solstice &	estimated	& 83.7	&	16.6 &	9.3 &	7.3 & 0.1	\\ \cline{2-7}
  &	reference	& 79.2 	&	20.8 &	6.3	 &	13.1 & 1.4 \\ \hline
  \multicolumn{2}{l}{actual (non-weighted)} &		71.4	&	28.6&	16.1		&9.3	&3.2 \\
\enddata
\end{deluxetable}

\subsection{Estimation of the Optical Depth of Atmosphere}
\label{s:tau}

So far we have assumed that the optical depth $\tau$ is indeed
equivalent to the input value $\tau _0$.  In reality, however, the value
of $\tau $ would not be known a priori, and thus $\tau $ should be regarded as
one of the fitting parameters.  The use of an incorrect value of $\tau $
would degrade the fit.  We investigate this issue by repeating the fit
with different input optical depths:
\begin{equation}
\label{eq:tau-n}
\tau _{\rm fit} = \frac{n}{10} \tau _0 \;\;\;\;\;(n=0,1,2, ... , 20) .
\end{equation}
The matrix $D_{jk}$ depends on the value of $\tau $
(Equations (\ref{eq:fatm}), (\ref{eq:C}), (\ref{eq:effalbedo}), and (\ref{eq:fmodel})).  
Assuming that the optical depth is constant over the planetary surface, we sum up 
Equation (\ref{eq:chi}) over different epochs according to 
\begin{equation}
\label{eq:sumchi}
\chi ^2 (\tau ) =  \displaystyle \sum _{i=1}^{24} \sum _{j=1}^5 
\frac{\left\{ I_{{\rm obs}\;j} (t_i) -  F_{*j} R_{\rm p}^2
\int _s \cos \theta _0 \cos \theta _1 ds \; 
\displaystyle \sum _k D_{jk}(\tau ) A_{k}(t_i) \right\} ^2}
{\sigma ^2 _{j}(t_i)},
\end{equation}
and search for the best-fit value of $\tau $ that minimizes Equation (\ref{eq:sumchi}) 
among the different values (Equation (\ref{eq:tau-n})). 

\begin{figure}[!tbh]
\centerline{\includegraphics[width=10.0cm]
{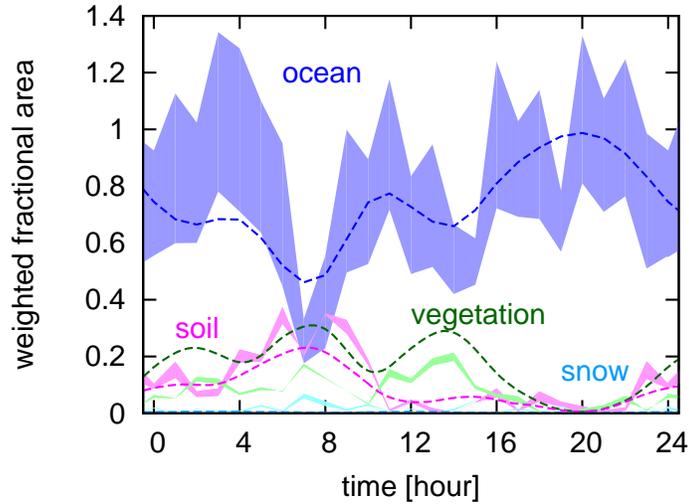}} \caption{Dependence of reconstructed fractional areas on the input 
value of $\tau _{\rm fit}$. The painted area exhibits the range of the estimated 
values when $\tau _{\rm fit} /\tau _0$ varies from 0.5 to 1.5. Simulated light 
curves in bands 1 to 5 are used.
The dashed lines show the weighted fractional areas derived from the MODIS land ecosystem 
classification dataset, like those in Figure \ref{fig:inv_scatter2}. 
}  \label{fig:taufit2}
\end{figure}

It would be instructive to see first how the reconstructed fractional areas 
are sensitive to the value of $\tau _{\rm fit}$. Figure
\ref{fig:taufit2} shows a case for which the simulated light curves
with $\tau=\tau_0$ are fit using
input values of $\tau _{\rm fit}$ varied from $0.5\tau_0$ to $1.5 \tau_0$. Since 
different values of $\tau _{\rm fit}$ mainly modify the effective albedo of ocean through
Rayleigh scattering, estimates of ocean fraction are sensitive to $\tau _{\rm fit}$; ocean 
area is overestimated (underestimated) for small (large) $\tau _{\rm fit}$. 
However, the estimate of the fractional areas for the other three surface types
is fairly robust, indicating that these fractional areas are determined mainly at 
longer wavelengths where the value of $\tau $ does not make any significant difference.

\begin{figure}[!tbh]
  \centerline{\includegraphics[width=10.0cm]{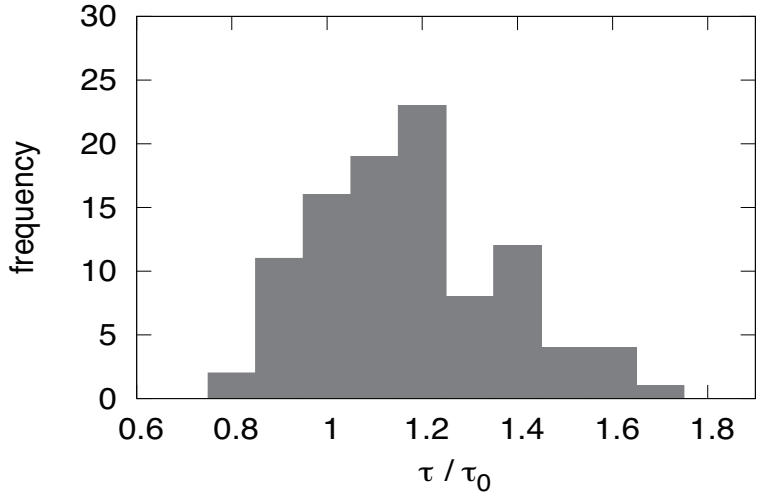}}
\caption{Histogram of best-fit values of $\tau _{\rm fit}$
from 100 realizations of simulated light curves in bands 1 to 5. }
  \label{fig:taufit}
\end{figure}

Figure \ref{fig:taufit} displays the histogram of the best-fit values of
$\tau _{\rm fit}$ (Equation (\ref{eq:tau-n})) which minimize Equation (\ref{eq:sumchi}).
The results show a broad peak around $\tau _{\rm fit} \sim 1.2\tau_0$, 
but cases of $\tau _{\rm fit} < 0.9\tau_0$ or $\tau _{\rm fit} > 1.6\tau_0$ are 
relatively rare. Given the crude approximations
adopted in our reconstruction method, these results are also encouraging.

Thus we have found that our inversion method can recover the presence of 
ocean and atmosphere simultaneously for a cloudless Earth-like planet. 
It may be also instructive to consider a hypothetical ocean planet
similar to Earth but without atmosphere at all.  In this case the ocean
contributes a tiny fraction of the total scattered light, and also
a small fraction in terms of area because of the specular reflection.

\begin{figure}[!tbh]
  \centerline{\includegraphics[width=12.0cm]
{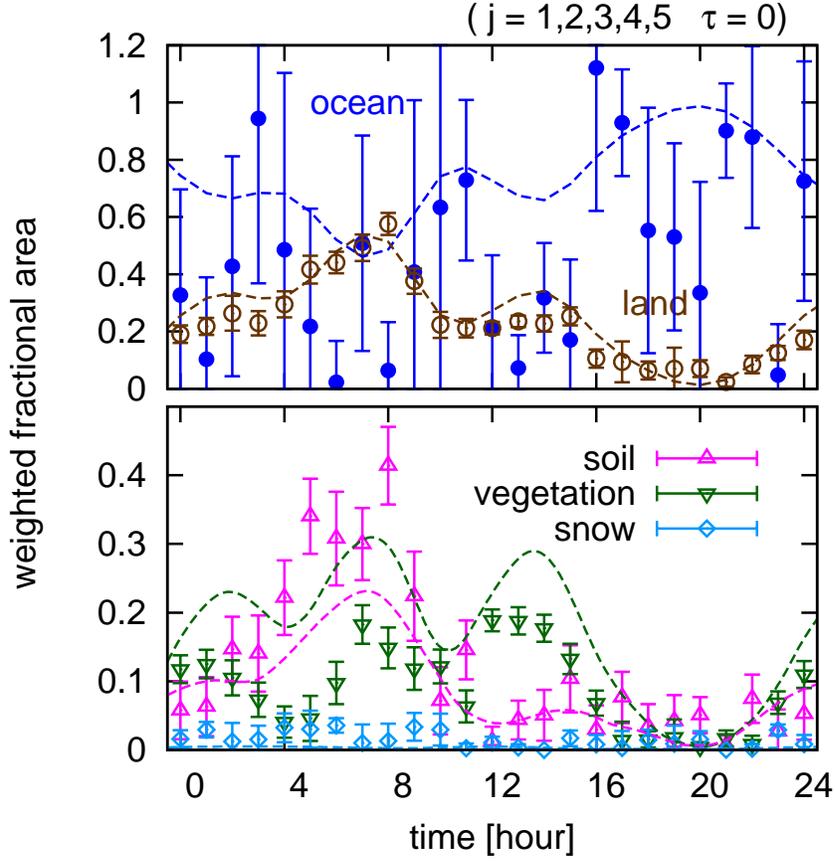}} 
\caption{Same as Figure \ref{fig:inv_scatter2} but without
an atmosphere.}  \label{fig:brdf3_noatmos}
\end{figure}

In order to see this more quantitatively, we create light curves for the
Earth without an atmosphere (i.e., Rayleigh scattering is neglected),
and repeat the same analysis.  The best-fit result with $\tau _{\rm fit} = 0$ in
Equation (\ref{eq:chi}) is shown in Figure \ref{fig:brdf3_noatmos}.  As
expected, the estimate of the fractional areas of ocean is very unstable and unreliable. 
It is interesting to note, however, that one can still
reconstruct the fractional areas of soil and vegetation fairly well. 

\section{Discussion}
\label{s:dis}

\subsection{Band Selection}
\label{s:band}

Our inversion method relies entirely on the difference of the
wavelength-dependence of the scattering spectra among ocean, soil,
vegetation, and snow, or their colors in short. 
Therefore the selection of the observed bands is crucial.

\begin{figure}[!tbh]
\centerline{\includegraphics[width=16.0cm]
{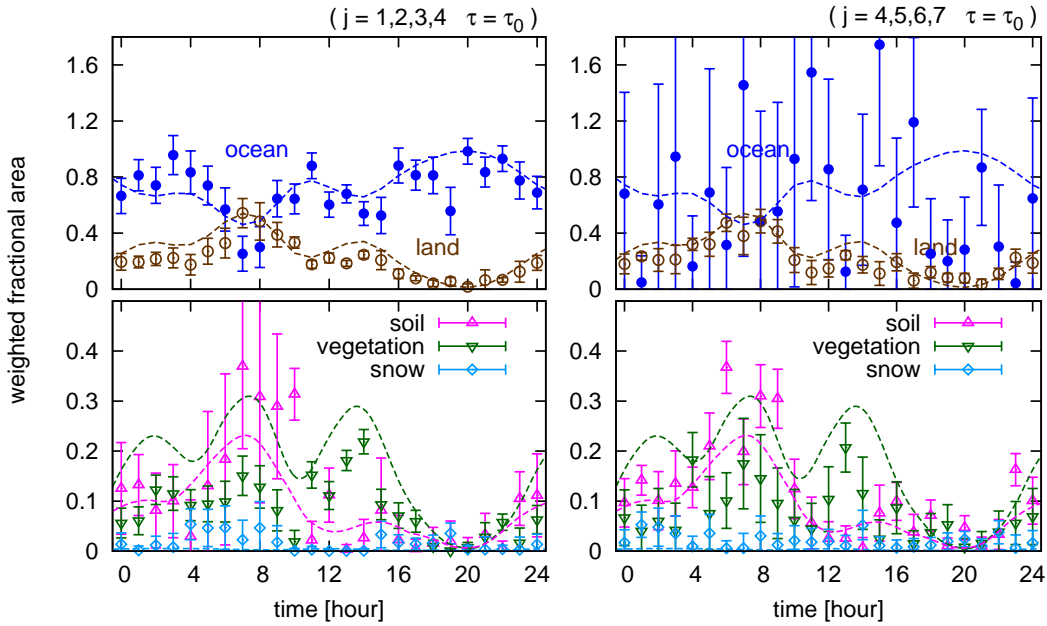}}
\caption{Reconstructed fractional areas using simulated light curves in 
two different sets of four bands; Left: bluer four bands ($j=$ 1, 2, 3, and 4), 
Right: redder four bands ($j=$ 4, 5, 6, and 7).}
  \label{fig:comparison2}
\end{figure}

First we repeat the same analysis performed in Section \ref{s:inverse} but with 
four bands.  In the case of five bands, we can fit up to five unknowns; we selected 
fractional areas for the four types and $\tau$ as the five fitting parameters. 
In the case of 4 bands, however, we cannot fit $\tau$ simultaneously and thus fix $\tau=\tau_0$.
The result is shown in Figure
\ref{fig:comparison2}.  The left and right panels correspond to the
bluer bands ($j=$ 1, 2, 3, and 4) and redder bands ($j=$ 4, 5, 6, and 7),
respectively. 
The difference can be easily understood; the bluer bands 
are more sensitive to the effect of Rayleigh scattering and the 
fractional area of ocean is recovered fairly well. Also the red edge feature
between bands 3 and 4 still carries the vegetation signature,
while the fractional area of soil becomes more uncertain because it is brighter 
in the redder bands. The result with the redder bands shows consistently opposite generic
features; the lack of the blue bands and the red edge makes it difficult
to detect the signature of ocean and vegetation, respectively, while the
soil is more easily reconstructed.

\begin{figure}[!tbh]
\centerline{\includegraphics[width=16.0cm]
{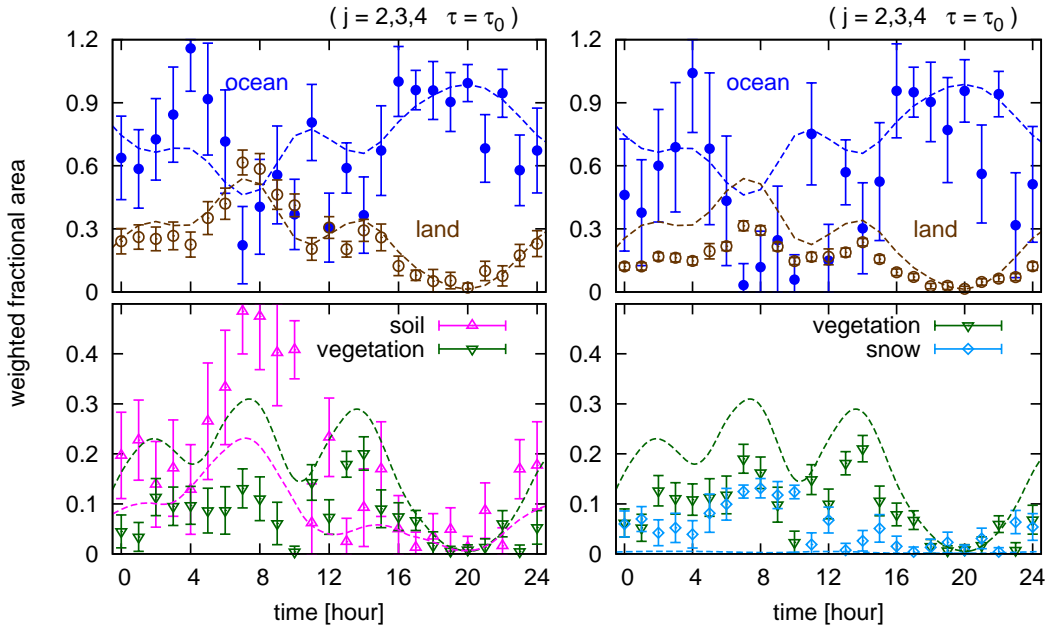}}
\caption{Reconstructed fractional areas using simulated light curves in 
three bands; Left: ocean, soil, and vegetation are considered. 
Right: ocean, vegetation, and snow are considered.}
\label{fig:band3}
\end{figure}

A closer look at Figure \ref{fig:comparison2} indicates some degeneracy between 
reflection spectra of selected surface types (Figure \ref{fig:effalbedo}). 
This is more clearly illustrated in Figure \ref{fig:band3}, where we use three 
bands ($j=2$, 3, and 4) only. In this case we cannot determine four surface
types, and we choose ocean, soil, and vegetation in the left panel, and ocean, 
vegetation, and snow in the right panel, while we neglect the remaining surface 
type from the fit. The result naturally is degraded compared with the 4-band 
and 5-band cases. The vegetation signature is still there because we chose bands 
3 and 4 that bracket the red edge, and the fractional areas of soil and snow 
compensate for each other. 
The fact that the spectra of snow and soil are similar in shorter
wavelength bands (except for their amplitude) partly explains the
behavior of the lower panels in Figure \ref{fig:band3}. While this similarity may
come largely from our single scattering approximation as described in
Section \ref{ss:inversion}, we obtained a similar result even when we use the snow
effective albedo of $\tau=0$. Therefore the degeneracy between snow
and soil is fairly generic as long as we use the short wavelength
bands in the reconstruction.

\subsection{Principal Component Analysis (PCA)}
\label{s:pca}

\begin{figure}[htbp]
  \begin{center}
   \includegraphics[width=12cm]{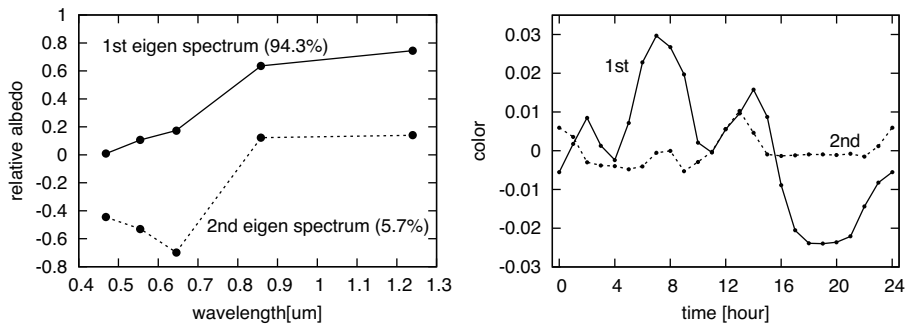}
\caption{
Left: the eigenspectra extracted by PCA of our fiducial mock light curves 
displayed in Figure \ref{fig:scatter2}. The contribution of the first eigenspectrum 
(solid) is 94.3 \% and that of the second one (dashed) is 5.7 \%.
Right: the time variation of the first eigenspectrum (solid) and the second 
eigenspectrum (dashed). 
} 
\label{fig:eigenvector5band}
\end{center}
\end{figure}

\citet{cowan2009} performed a PCA of the multi-band light
curves of the Earth as observed by the EPOXI mission.  They extracted two
major eigenspectra in a model-independent manner.
We also performed PCA on the same data-set, and obtained results consistent 
with those of \citet{cowan2009}. 
We then performed PCA of our mock light curves and again extracted two major 
eigenspectra; these are shown in Figure \ref{fig:eigenvector5band}. 

\begin{figure}[htbp]
  \begin{center}
   \includegraphics[width=12cm]{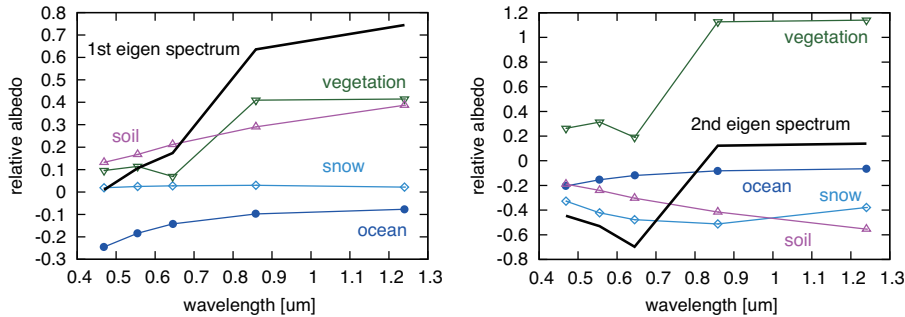}
\caption{
Left: decomposition of the first eigenspectrum of Figure \ref{fig:eigenvector5band} 
where the coefficients are ocean, -2.19; snow, 0.03; vegetation, 0.86; soil, 1.07. 
Effective albedos with $\tau = \tau _0 $ are used. Right: decomposition of the 
second eigenspectrum of Figure \ref{fig:eigenvector5band} where the coefficients are 
ocean, -1.83; snow, -0.58; vegetation, 2.36; soil, -1.54. Effective albedos with 
$\tau = \tau _0 $ are used.
}
\label{fig:pcadecompose}
\end{center}
\end{figure}

In order to compare our current methodology with PCA, we decompose these
two eigenspectra into a combination of the effective albedo of the four surface 
types with $\tau = \tau_0$ (solid lines in Figure \ref{fig:effalbedo}).  
We find that the first
eigenspectrum roughly corresponds to (soil+vegetation-ocean), and the
second one roughly corresponds to (vegetation-(soil+snow+ocean)) as
displayed in Figure \ref{fig:pcadecompose}.  

This result indicates that
the extracted eigenspectra do not necessarily correspond to any single
surface type.  This is not surprising, of course.  
While PCA extracts orthogonal eigenspectra by definition, the
wavelength dependence of albedos of real materials is not orthogonal in general.
Moreover, they are likely to be degenerate in PCA because the fractional
areas are complementary; for example, the fractional area of ocean
decreases when fractional area of land increases. 
In other words, the time dependences of these components are necessarily correlated.
Therefore, it seems natural that the first eigenspectrum is
roughly (land-ocean).  
Our method is quite model dependent, but it allows
decomposition of the light curves into physical components
that can be interpreted in a simple way. 
While the model independence of PCA is a great advantage, 
the final interpretation is not straightforward.
Further comparison with PCA is beyond the scope of this paper, 
but clearly these two methodologies are very complementary.

\section{Summary}
\label{s:sum}
 
In this paper we have presented a method to reconstruct the fractional areas of 
different surface components from photometric colors of Earth-like exoplanets without 
clouds. For this purpose we first created mock light curves in seven photometric bands 
from the observed data of the Earth, but neglected the  clouds. The light curves 
are fitted to isotropic scattering models consisting of four surface types: ocean, 
soil, snow, and vegetation. In a very idealized situation where the data are obtained 
using a noiseless 2 m telescope and multiple integrations of 1 hr each, we find that 
our method is able to reproduce the fractional areas of those components fairly well. 
In particular, Figures \ref{fig:inv_scatter2}, \ref{fig:taufit2}, \ref{fig:brdf3_noatmos}, 
\ref{fig:comparison2}, and \ref{fig:band3} show quantitatively that the presence of 
vegetation can be recovered using the color information via its red edge feature. 
Although we fit the fractional areas of each component independently at each time 
in the light curves, the time variation is eventually translated into the distribution 
along the longitude with the methodology described in \citet{cowan2008}.

We also find that Rayleigh scattering due to the atmosphere plays a key role in 
estimating reliably the fractional areas of ocean.
Indeed, without an atmosphere, our method based on the isotropic scattering assumption 
cannot properly estimate the real fractional area of ocean because of the strongly 
anisotropic nature of its specular reflection. On the other hand, for terrestrial 
exoplanets with atmosphere similar to our Earth, we may be able to estimate the 
presence of ocean and atmosphere simultaneously if the effect of clouds is safely neglected.

Our methodology described in this paper is based on admittedly several
very idealized assumptions and simplifications. There are a variety of
issues that we have to address and improve in future work, some of which are listed below.  

First, one of the most serious omissions in the present modeling is the
absence of clouds. 
Clouds provide additional time variation in the
light curves, which is not directly related to the property of the
planetary surface.  Clouds greatly increase the reflectivity in
visible and near-infrared bands.  We may treat clouds as an additional
component in our current model. While the time and spatial variabilities of clouds 
are not easy to model, combining data from multiple rotation periods may allow us to 
separate variations due to surface features from those caused by clouds. 
We will discuss these effects of clouds more quantitatively in future work.

Second, the universality of the scattering spectra of surface types on
Earth is not clear, maybe not even likely. Indeed this could be regarded
as both the strongest and the weakest point in our methodology; we can recover 
surface information, including the presence of vegetation from the color 
variations alone because our spectral template is derived from that of 
the Earth. Nevertheless, as stressed in the introduction, it is an important 
first attempt to see if we could infer the presence of vegetation 
observationally for exoplanets similar to the Earth even in the most favorable 
and idealized circumstances. The answer to the question seems promising and 
the next step is to generalize the result for a wider range of possibilities. 
The scattering properties of ocean and pure snow may be universal because 
the properties of ${\rm H}_2{\rm O}$ would not be different on different planets, 
but they can be somewhat different depending on impurities, for example. 
The generality of the red edge depends on the photosynthesis system of 
exo-vegetation and the wavelength of that ``edge'' is likely to shift according 
to the spectral type of the host star \citep{wolstencroft2002, kiang2007a, kiang2007b}. 
Also, the scattering properties of soil may vary depending on the details 
of the soil's composition, and even the gravity may affect the granularity 
which changes the scattering properties.  

Finally we can explore other geometrical configurations of a
star-planet system including the effects of orbital inclination, planet 
obliquity, and seasonal variations. 
It will be interesting to
apply our methodology to time-series remote sensing data-sets for the Earth 
as well as to models of the Earth in the distant past which take into account 
continental drift and the evolutionary history of vegetation.

\acknowledgments 

Y.F. and Y.S. gratefully acknowledge support from the Global
Collaborative Research Fund (GCRF) 
``A World-wide Investigation of Other Worlds'' grant
and the Global Scholars Program of Princeton
University, respectively. 
They also thank people in Peyton Hall, 
Princeton University, where this paper was completed, for their warm  
hospitality and discussions. 
In particular, we are grateful to Adam Burrows for his helpful comments on this paper. 
H.K. is supported by JSPS (Japan Society for the Promotion of Science) Fellowship 
for Research (DC2:20-10466), A.T. by a JSPS grant (No.21740168).  E.L.T is also 
supported in part by the aforesaid GCRF grant and the World Premier International 
Research Center Initiative (WPI Initiative), Ministry of Education, Culture, 
Sports, Science and Technology, Japan. This work is also supported by
JSPS Core-to-Core Program ``International Research Network for Dark Energy.''

{\appendix

\section{BRDF for land \label{ap:landBRDF}}

We create mock light curves adopting the Rossi-Li model
(Equation (\ref{eq:Rossi-Li})) for the land BRDF. 
The specific
expressions for the volume-scattering term, $K_{{\rm vol}}$, and the
geometric-optical term, $K_{{\rm geo}}$, are given here for
completeness.  The derivation of the two terms is found in
\citet{wanner1995}. 
The volume-scattering term is
\begin{eqnarray}
K_{{\rm vol}}(\theta _0 , \theta _1 , \phi ) = 
\frac{( \pi / 2 - \xi ) + \sin \xi}{\cos \theta _0 + \cos \theta _1} 
- \frac{\pi }{4}, 
\end{eqnarray}
where
\begin{eqnarray}
\cos \xi = \cos \theta _0 \cos \theta _1 
+ \sin \theta _0 \sin \theta_1 \cos \phi .
\end{eqnarray}

The geometric-optical term is:
\begin{eqnarray}
K_{{\rm geo}}(\theta _0 , \theta _1 , \phi ) =
O(\theta _0 , \theta _1, \phi ) - \sec \theta ' _0  - \sec \theta ' _1 
+ \frac{1}{2}(1 + \cos \xi ') \sec \theta' _0 \sec \theta'_1,
\end{eqnarray}
where
\begin{eqnarray}
O(\theta _0, \theta _1, \phi ) 
&=& \frac{1}{\pi}(t - \sin t \cos t)(\sec \theta ' _0 + \sec \theta ' _1), \\
\cos t &=& {\rm min} \left\{ 1, \frac{h}{b} 
\frac{\sqrt{D^2 + (\tan \theta ' _0 \tan \theta ' _1 \sin \phi )^2}}
{\sec \theta ' _0 + \sec \theta ' _1} \right\}, \\
D &=& \sqrt{\tan ^2 \theta ' _0 + \tan ^2 \theta '_1 
- 2 \tan \theta' _0 \tan \theta ' _1 \cos \phi }, \\
\cos \xi ' &=& \cos \theta '_0 \cos \theta '_1 
+\sin \theta ' _0 \sin \theta '_1 \cos \phi, \\
\theta ' _0 &=& \tan ^{-1} \left( \frac{b}{r} \tan \theta _0 \right) , \\
\theta '_1 &=& \tan ^{-1} \left( \frac{b}{r} \tan \theta _1 \right) .
\end{eqnarray}
This kernel assumes a sparse ensemble of surface objects which throws shadows on the Lambertian background.  
The objects are approximated as spheroids with width $2r$ and vertical length $2b$. The height of the center of the spheroids is denoted by $h$. 
The function $O(\theta_0, \theta _1, \phi ) $ is the overlap area between the solar shadows and the shade of view. 
For MODIS processing $h/b=2$ and $b/r = 1$ are assumed (i.e., the spherical crowns are separated from the ground by their radii).  Thus, we adopted these values.

\section{BRDF for ocean \label{ap:oceanBRDF}}

The BRDF model for ocean is very complicated, and we summarize the key
expressions of the model by \citet{nakajima1983} that we adopt in the
present work.

The scattering of solar radiation by a flat ocean follows the simple
Snell-Fresnel law.  Using the pair $\{\theta ,\phi \}$ to represent
polar coordinates relative to the normal of the averaged surface plane
(Figure \ref{fig:config_ocean}), the relation between the incident
radiance $I_{\rm in}(\theta ' , \phi ')$ and the scattered radiance
$I_{\rm out}(\theta , \phi) $ is written as
\begin{eqnarray}
\label{eq:NT-I}
I_{{\rm out}} (\theta , \phi ) &=& 
\int d\cos \theta ' d\phi ' R(\theta ,\theta ' ,\phi-\phi ') 
I_{{\rm in}} (\theta ', \phi '), \\
\label{eq:NT-R}
R(\theta ,\theta ' ,\phi-\phi ') &=& r(\theta , \tilde m) 
\delta(\cos\theta - \cos \theta ') \delta (\phi ' - \phi - \pi ),
\end{eqnarray}
where $r(\theta , \tilde m) $ is the Fresnel scattering coefficient
and $\tilde m$ the is the ratio of the reflective indices of air and ocean.
When the wind above ocean is taken into account, however, the slope of
scattering surface varies both temporally and spatially. 

\begin{figure}[!tbh]
  \centerline{\includegraphics[width=6.0cm]{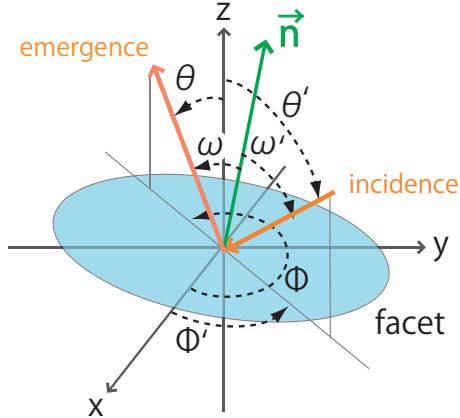}}
  \caption{Configuration of incident and scattered rays with respect to
  a wave facet whose normal vector ${\bf n}$ is \{$\theta_n$,
  $\phi_n$\} in the polar coordinates with respect to the average surface
  plane.}  \label{fig:config_ocean}
\end{figure}
Thus, the normal direction of the facet is different from that of the
averaged surface plane ($z$-axis in Figure \ref{fig:config_ocean}).  We
introduce the polar coordinate $\{\omega , \xi \}$ with respect to the
normal vector ${\bf n}$ of each facet as shown in Figure
\ref{fig:config_ocean}. Then the scattered radiance including the effect of rough 
surface is written as
\begin{eqnarray}
I_{{\rm out}} (\theta ,  \phi ) 
&=& \frac{1}{\cos \theta }\int d\cos \theta _n 
\int d \phi _n \int d\cos \theta ' 
\int d\phi ' \cos \omega  \cr
&& \qquad \times S(\theta , \theta ', \phi-\phi ', \theta _n,\phi _n) 
R(\omega , \omega ' ,\xi-\xi ') I_{{\rm in}} (\theta ', \phi ').
\end{eqnarray}
In the above, $\{\theta _n,\phi _n \}$ denotes the direction of the
normal of the facet, and $S $ is the effective fractional area of the wave
facet associated with $\{ \theta _n , \phi _n \}$.  The integration over
$\theta _n$ and $\phi_n$ results in setting $\omega = \omega'$ due to
the Dirac delta in Equation (\ref{eq:NT-R}). Then Equation
(\ref{eq:NT-I}) reduces to
\begin{eqnarray}
I_{{\rm out}} (\theta ,  \phi ) 
&=& \int d\cos \theta ' \int d \phi ' 
\left| \frac{\partial (\cos \theta _n, \phi _n)}
{\partial (\cos \theta ', \phi')}\right| 
\cos \omega ^* \cr
&& \qquad \times S(\theta , \theta ', \phi-\phi', \theta _n^* , \phi _n^*) 
r(\omega^* , \tilde m ) I_{{\rm in}} (\theta ' , \phi '),
\label{eq:oceanref}
\end{eqnarray}
where $\theta _n^*$, $\phi _n^*$, and $\omega ^*$ are defined through the
conditions of $\omega = \omega '$ and $\xi - \xi' = \pi$, or more
explicitly,
\begin{eqnarray}
\cos \theta _n^* 
&=& \frac{\cos \theta + \cos \theta '}{2 \cos \omega ^*} 
\label{eq:normal} \\
\cos (2\omega ^*) &=& \cos \theta \cos \theta ' + \sin \theta \sin \theta ' \cos (\phi - \phi ')
\label{eq:omega}
\end{eqnarray}
as shown in Figure \ref{fig:config_ocean}.
The Jacobian determinant in Equation (\ref{eq:oceanref}) is
\begin{equation}
\left| \frac{\partial (\cos \theta _n, \phi _n)}{\partial (\cos \theta ', \phi ')} \right| = \frac{1}{4 \cos \omega ^*}
\label{eq:jacobian}
\end{equation}

The effective fractional area $S $ is expressed as
\begin{equation}
\Sigma S d\cos \theta _n d \phi _n 
= \frac{\Sigma}{\cos \theta _n} P(\theta _n, \phi _n ) 
\mathcal{T}(\theta _0, \theta _1 '; \phi, \phi ' | \theta _n , \phi _n ) 
d\cos \theta _n d \phi _n ,
\end{equation}
where $\Sigma $ is the horizontal area, $P$ is the density function of
the wave slope, and $\mathcal{T} $ is the bidirectional shadowing factor. 
The factor $P(\theta _n , \phi _n )$ is empirically given
\citep[see Figure 3 of][]{nakajima1983} by
\begin{equation}
\label{eq:thetan-pdf}
P(\theta _n , \phi _n) 
= \frac{1}{\pi \sigma ^2 \cos^3 \theta _n}
\exp\left( -\frac{\tan^2 \theta _n}{\sigma ^2} \right) \equiv p (\theta _n).
\end{equation}
Here, $\sigma $ is the rms of slopes and a function of the wind velocity, $u_{10}$, at $10\;\;{\rm m}$ height above the water surface taken to be
\begin{equation}
\label{eq:oceansigma}
\sigma ^2 = 0.00534 u_{10}.
\end{equation}
Another factor, $\mathcal{T} (\theta , \theta ' ; \phi , \phi '; \theta
_n , \phi _n)$, is obtained from an analytical fit to the results of
a Monte Carlo simulation assuming one-dimensional random surface with
Gaussian auto-correlation, which resulted in
\begin{eqnarray}
\mathcal{T}(\theta, \theta ' ; \phi , \phi '| \theta _n , \phi _n) 
&=&H\left( v - \frac{\partial z / \partial x}{\sigma } \right) 
H\left( v' - \frac{\partial z / \partial x}{\sigma } \right) G(v, v'), \\
v &=& \frac{\cos \theta }{\sigma \sin \theta}, \;\; v' 
= \frac{\cos \theta ' }{\sigma \sin \theta '}, \\
G(v, v') &=& \frac{1}{1 + F(v) + F(v')}, \\
F(v) &=& \displaystyle \frac{1}{2}
\left[ \frac{\exp (-v^2)}{\sqrt{\pi} v^2} 
- \frac{2}{\sqrt{\pi}}\int _v ^{\infty} \exp ( - t^2) dt \right],
\end{eqnarray}
where $\partial z / \partial x$ is the slope of the facet, $H(x)$ is the Heaviside step function, and $\sigma $ is given by
Equation (\ref{eq:oceansigma}).

Combining all the expressions above, Equation (\ref{eq:oceanref}) is now
written as:
\begin{eqnarray}
\label{eq:oceanref3}
I_{\rm out}(\theta , \phi )&=& \int d\cos \theta ' 
\int d \phi R(\theta , \theta ', \phi-\phi ') 
I_{\rm in}(\theta ', \phi'), \\
R(\theta , \theta ', \phi - \phi') 
&=& \frac{1}{\cos \theta \cos \theta _n^*} 
p(\theta _n^*) G(v , v') r(\omega ^* , \tilde m),
\end{eqnarray}
where $\theta _n ^*$ and $\omega ^* $ are functions of $\theta ,\theta '
$, and $(\phi -\phi' )$ through Equations (\ref{eq:normal}) and
(\ref{eq:omega}). 

Considering a parallel incident flux $F_*$ from the direction of $\{
\theta _0, \phi _0 \}$, we have
\begin{equation}
\label{eq:oceanref4}
I_{\rm in}(\theta , \phi) 
= F_* \delta (\cos \theta ' - \cos \theta _0) \delta(\phi'- \phi_0).
\end{equation}
Substituting Equation (\ref{eq:oceanref4}) into Equation
(\ref{eq:oceanref3}) and replacing $\{ \theta ,\phi \}$ with $\{\theta _1,
\phi_1\}$, we finally arrive at
\begin{eqnarray}
I_{\rm out}(\theta _1, \phi _1 )
&=& F_* R(\theta _0 , \theta _1, \phi_0-\phi_1), \\
R(\theta _0, \theta _1, \phi_1 - \phi_0) &=& 
\frac{1}{\cos \theta_1 \cos \theta _n^*} 
p(\theta _n^*) G(v , v') r(\omega ^* , \tilde m) .
\end{eqnarray}
This is equivalent to Equation (\ref{eq:oceanBRDF}) in
Section \ref{subsubsec:oceanBRDF} :
\begin{equation}
f_{\rm ocean}(\theta _0, \theta _1, \phi) 
= \frac{1}{4 \cos \theta _0 \cos \theta _1  \cos \theta _n^*} 
p(\theta _n^*)G(v, v')r(\omega ^*, \tilde m).
\end{equation}



\end{document}